# The Mid-Infrared Colours of Galactic Bulge, Disk and Magellanic Planetary Nebulae

J.P. Phillips[1] and G. Ramos-Larios[1,2]

[1] Instituto de Astronomía y Meteorología, Av. Vallarta No. 2602, Col. Arcos Vallarta, C.P. 44130 Guadalajara, Jalisco, México   e-mail : jpp@astro.iam.udg.mx

Current address:
[2] Instituto de Astrofísica de Andalucía, IAA-CSIC
C/ Camino Bajo de Huétor # 50, 18008,
Granada, Spain   e-mail : gerardo@iaa.es

**Abstract**

We present mid-infrared (MIR) photometry for 367 Galactic disk, bulge and Large Magellanic Cloud (LMC) planetary nebulae, determined using GLIMPSE II and SAGE data acquired using the Spitzer Space Telescope. This has permitted us to make a comparison between the luminosity functions of bulge and LMC planetary nebulae, and between the MIR colours of all three categories of source. It is determined that whilst the 3.6 $\mu$m luminosity function of the LMC and bulge sources are likely to be closely similar, the [3.6]-[5.8] and [5.8]-[8-0] indices of LMC nebulae are different from those of their disk and bulge counterparts. This may arise because of enhanced 6.2 $\mu$m PAH emission within the LMC sources, and/or as a result of differences between the spectra of LMC PNe and those of their Galactic counterparts. We also determine that the more evolved disk sources listed in the MASH catalogues of Parker et al. and Miszalski et al. (2008) have similar colours to those of the less evolved (and higher surface brightness) sources in the catalogue of Acker et al. (1992); a result which appears at variance with previous studies of these sources.

# 1. Introduction

The MIR fluxes of planetary nebulae (PNe) depend upon a broad range of emission mechanisms, including warm dust continuum emission, emission associated with PAH bands and plateau features, ionic and atomic permitted and forbidden line transitions, the quadrupole transitions of $H_2$, and the underlying free-free and bound-free components of gaseous continuum emission. Given that at least certain of these components depend upon the degree of excitation of the gas, the mass and evolutionary status of the shells, the presence or otherwise of shock and fluorescent excitation, and the chemical composition of the progenitors, it follows that MIR spectra may vary appreciably between differing nebular outflows.

In particular, we note that the abundances of Galactic bulge PNe (GBPNe) differ from those of Galactic disk sources (GDPNe), tending as they do to have slightly higher metallicities (e.g. Wang & Liu 2007; Maciel et al. 2006; Gorny et al. 2004; although see also Exter et al. (2004) and Casassus et al. (2001), where it is noted that O/H ratios are indistinguishable within rather large error limits), lower C/O ratios (Wang & Liu 2007; Casassus et al. 2001), and differing abundance trends for other elements indicative of markedly different evolutionary histories (see e.g. Cuisinier et al. 2002). It has also been noted that the sources are likely to possess differing radial trends in abundance gradient; that bulge sources have a larger incidence of oxygen rich features, such as the crystalline silicate bands at ~ 26-36 $\mu$m (Perea-Calderón et al. 2009; Gutenkunst et al. 2008; Casassus et al. 2001); and that a larger fraction of bulge sources sport both oxygen and carbon rich features in the MIR (Gutenkunst et al. 2008). Indeed, such mixed features appear to be present in all of the 26 sources investigated by Perea-Calderón et al. (2009).

The GDPNe are noted for radial gradients in abundance, with those located close to the Galactic centre having higher metallicities (see e.g. Martins & Viegas (2000); Maciel & Quireza (1999), Maciel & Koppen (1994); Pasquali & Perinotto (1993); Amnuel (1993); Samland et al. (1992); Koppen et al. (1991); and Faundez-Abans & Maciel (1986)). This is attributed to the synthesis of elements over the lifetime of the Galaxy, and the varying degrees to which they enrich the interstellar medium, and affect the abundances of the progenitors.

Finally, and at the other extreme, it has been noted that PNe within the Magellanic Clouds have metallicities which are between ~0.25 dex (LMC) and ~0.5 dex (SMC) lower than those of GDPNe (Maciel et al. 2006, 2008; Chiappini et al. 2008); a disparity which may lead to differences in ionising fluxes, central star luminosities, emission line strengths, and grain formation and emission characteristics.

It is therefore of interest to explore such influences upon the MIR characteristics of the nebulae, a program which may be undertaken through either spectroscopy [see e.g. Gutenkunst et al. (2008) and Perea-Calderón (2009) for bulge nebulae, Bernard-Salas et al. (2005) and Stanghellini et al. (2007) for LMC and SMC PNe, and Hora et al. (2004) and Phillips & Ramos Larios (2008a) for GDPNe] or broad-band photometry and mapping [see e.g. Hora et al. (2008) for LMC sources, and Phillips & Ramos-Larios (2008a, b); Ramos-Larios & Phillips (2008a, b); Hora et al. (2004); Cohen et al. (2007a); Su et al.(2004, 2007); and Ueta (2006) for GDPNe].

We shall, in the following, be adding to these latter data bases, and investigating the MIR colours of GBPNe contained within the catalogue of Van de Steene & Jacoby (2001); of LMC nebulae in the catalogue of Reid & Parker (2006); and of disk PNe listed in the catalogues of Acker et. al. (1992), Parker et al. (2006; MASH I) and Miszalski et al. (2008; MASH II). These latter catalogues refer to nebulae having a large range of evolutionary states, from the predominantly less evolved and higher surface brightness sources of Acker et al. (1992), to the larger, more evolved, and predominantly lower surface brightness nebulae of the two MASH catalogues.

We shall compare the colours and magnitudes of these disparate groups of nebulae, and note a broad range of similarities and a variety of differences. We shall also note a difference between the characteristics of our present MASH nebulae, and the results of previous analyses of these sources.

## 2. The Observational Data Base

We have acquired photometry and mapping of a wide range of Galactic disk, Galactic bulge, and LMC planetary nebulae using data deriving from the second Galactic Legacy Infrared Midplane Survey Extraordinaire (GLIMPSE II) (Benjamin et al. 2003), and the program Surveying the Agents of a Galaxy's Evolution (SAGE; Meixner et al. 2006). Both of these surveys were undertaken using the Spitzer Space Telescope (SST; Werner et al. 2004). Details of the instrumentation and data processing for these programs have been described in previous publications (see e.g. Hora et al. 2004, 2008; Phillips et al. 2008a, b), and we shall note only the most salient features of importance to this analysis.

The catalogues and mosaics of the GLIMPSE II project were published over a period extending between 2007 and the Spring of 2008, and cover approximately 25 square degrees of the Galactic plane within the regimes 350° < $\ell$ < 10°, and -2° < b < +2°. The mapping was undertaken using the Infrared Array Camera (IRAC; Fazio et al. 2004), and employed filters having isophotal wavelengths (and bandwidths $\Delta\lambda$) of 3.550 μm ($\Delta\lambda$ = 0.75 μm), 4.493 μm ($\Delta\lambda$ = 1.9015 μm), 5.731 μm ($\Delta\lambda$ = 1.425 μm) and 7.872 μm ($\Delta\lambda$ = 2.905 μm). The spatial resolution varied between ~1.7 and ~2 arcsec (Fazio et al. 2004), and is reasonably similar in all of the bands, although there is a stronger diffraction halo at 8 μm than in the other IRAC bands. This leads to differences between the point source functions (PSFs) at ~0.1 peak flux. The maps were published at a resolution of 0.6 arcsec/pixel.

Further details of the GLIMPSE data processing are provided in http://www.astro.wisc.edu/sirtf/glimpse1_dataprod_v2.0.pdf.

We have used these recent GLIMPSE II results to undertake photometry and mapping of PNe close to the projected location of the Galactic centre. The positional listing of the nebulae derives from four principal catalogues. In the first place, the Acker et al. (1992) catalogue of Galactic PNe contains all of the sources identified prior to 1992. Most of these have reasonably high surface brightnesses, and correspond to less evolved nebular outflows, although there is also a sprinkling of larger, fainter, and (usually) higher Galactic latitude sources which correspond to later phases of nebular evolution. The fraction of the latter, more evolved sources is relatively small, however, and is expected to be even smaller within our present Galactic centre sample.

More recently, this list has also been supplemented by the so-called MASH I and MASH II catalogues of Parker et al. (2006) and Miszalski et al. (2008). These sources were detected as a result of investigation of the Anglo-Australian Observatory UK Schmidt Telescope (AAO/UKST) H$\alpha$ survey of the southern Galactic plane (Parker et al. 2005), and represent outflows which are mostly larger, fainter, and correspond to more evolved nebular shells.

Insofar as we can make a judgement based upon H$\beta$ or radio fluxes, and the angular sizes of the nebular shells, it appears that few if any of the present MASH and Acker et al. sources correspond to GBPNe. We shall be assuming that the large majority of these nebulae correspond to foreground disk PNe – many possibly located in the inner Galactic disk, where metallicities are higher.

The fluxes of the sources have been measured using a variety of circular or elliptical apertures, and employing GLIMPSE II FITS pipeline images. Backgrounds were estimated for nearby regions of sky, and using the same aperture sizes and shapes as were used to determine the total nebular fluxes. These were then subtracted from the nebular results, and the fluxes converted to magnitudes using the Vega calibration of Reach et al. (2005).

Whilst such results are entirely reasonable where source sizes are small (< 8-9 arcsec), scattering of light in the array focal plane affects larger aperture measurements – an effect which is particularly important at 5.0 $\mu$m and 8.0 $\mu$m, and may arise due to scattering in an epoxy layer between the detector and multiplexer (Cohen et al. 2007b). It is therefore necessary to modify fluxes as described in Table 5.7 of the SST IRAC handbook (see e.g. http://ssc.spitzer.caltech.edu/irac/dh/iracdatahandbook_3.0.pdf), and this leads to corrections of maximum order 0.944 at 3.6 $\mu$m, 0.937 at 4.5 $\mu$m, 0.772 at 5.8 $\mu$m and 0.737 at 8.0 $\mu$m. We have applied these changes for the larger sources considered here. However, the precise values of these corrections also depend on the underlying surface brightness distribution of the source, and for objects with size ~several arcminutes it is counselled to use corrections which are somewhat smaller (i.e. between the values cited above and unity). The handbook concludes that "this remains one of the largest outstanding calibration problems of IRAC".

It is therefore possible that some of the photometry for the larger sources may have been over-corrected, although errors are unlikely to exceed a tenth of a magnitude or so – and are certainly insufficient to radically affect the analysis below.

Finally, errors in these fluxes tend to be dominated by uncertainties in sky and nebular subtraction. In particular, sky background fluxes may be afflicted by weak stellar emission, sometimes barely discernable in images of the fields; variable levels of diffuse Galactic emission (particularly important at wavelengths $\geq 8$ μm); and components arising from photon and instrumental noise. We have determined these contributions in assessing the errors quoted in Tables 1 & 2, estimating the most important of these (background subtraction) by measuring the sky at a variety of locations about the sources.

The corresponding photometry for 26 Acker et al. (1992), and 20 MASH I & II sources are listed in Table 1, where we specify the Galactic coordinates of the sources (column 1), the source catalogue (column 2); the name of the source (column 3), and 3.6 μm to 8.0 μm photometry and associated errors (columns 4 to 11).

Some of the first identifications of Galactic Bulge PNe were undertaken by Pottasch & Acker (1989), selected on the basis of their proximity to the Galactic centre, small angular dimensions and low H$\beta$ flux levels. Subsequent listings of these sources have been provided by Beaulieu et al. (1990) and Van de Steene & Jacoby (2001), whilst a catalogue of a further 400 sources has been described by Acker et al. (2006). Most of these listings are either unavailable at present (viz. the case of the Acker et al. catalogue), or refer to nebulae which are, for the most part, located outside of the limits of the GLIMPSE II survey. An exception in this respect is the survey of Van de Steene & Jacoby (2001), in which the authors identified 64 sources in the radio continuum at 3 and 6 cm. We have determined IRAC photometry for approximately half of these sources, using data deriving from the GLIMPSE II PSC. For the remaining cases, we have determined MIR photometry from GLIMPSE imaging results, and using the procedures described above; a strategy which is necessary where the sources are appreciably extended, or have fluxes falling below the inclusion limits of the PSC. Photometry for these sources is listed in Table 2, where the sequencing of the columns is similar to that in Table 1.

Finally, Hora et al. (2008) have published equivalent photometry for nebulae within the LMC, making use of data products deriving from the Spitzer SAGE program (Meixner et al. 2006). In this case, the IRAC instrument, and the Multiband Imaging Photometer for Spitzer (MIPS; Reike et al. 2004) were used to perform an unbiased ~7°x7° survey centred on the LMC. The catalogue of Hora et al. contains some 275 sources, and makes use of the listing provided by Leisy et al. (1997). Reid & Parker (2006) have subsequently published positions for a further 460 new possible, likely or true PNe, however, and we have determined Spitzer and 2MASS magnitudes for these sources using the SAGE PSC. This catalogue, corresponding to the version published in Winter of 2008, is described as being "more reliable" than previous versions of the document; it includes sources having a S/N > 10 at 8 $\mu$m, and > 6 in the other wavebands, and excludes near-neighbour detections within a radius of 2 arcsec of the sources. It is clear however that not all of the gremlins have been successfully removed, and there is some duplication of photometry for certain of the sources.

We have identified MIR sources within 3 arcsec of the positions listed by Reid & Parker (2006), and assumed that the PNe correspond to the MIR detections with smallest positional offsets. The mean positional displacement is found to be 1.0 arcsec.

Details of the photometry are listed in Table 3, with the first column giving the number of the source in the catalogue of Reid & Parker, followed by the Galactic position (columns 2 and 3), the name and status of the source (columns 4 & 5; where T $\equiv$ True, P $\equiv$ Possible, and L $\equiv$ Likely), and photometry and associated errors (columns 6-19). The final column gives the displacement between the SAGE source position and the catalogue coordinates of Reid & Parker (2006).

We shall, for the purposes of the following analysis, fold this photometry in with the results of Hora et al. (2008), yielding a sample of some 178 nebulae for which photometry is available in all four of the IRAC bands. These results will subsequently be analysed within colour-colour and colour-magnitude planes, as described in Sects. 3 & 4.

## 3. The MIR Colour-Magnitude Diagrams

Hora et al. (2008) have undertaken an MIR colour-magnitude analysis for PNe located in the LMC and Galactic disk, taking account of the differing distance moduli of the respective groups of sources. Although such an analysis is certainly of interest, and leads to some intriguing results, it relies upon the extremely uncertain distances to the GDPNe. It would be more useful, under these circumstances, to compare the relative distributions of sources for which distances to the nebulae are reasonably well established.

We have undertaken such an analysis in Figs. 1 and 2, where we compare the distribution of LMC PNe (indicated by the smaller blue disks) with those of the GBPNe (larger red disks). In this case, the magnitudes of the LMC nebulae have been modified so as to correspond to the values they would have at the Galactic centre; where the distance to the LMC is assumed to be 48.1 kpc (Macri et al. 2006), and that of the Galactic centre is $7.62 \pm 0.32$ kpc (Eisenhauer et al. 2005). Subsequent to undertaking this analysis, we noted the publication, by the same team of authors, of an upgraded distance $8.33 \pm 0.35$ kpc (Gillessen et al. 2008), suggesting that the distance to the centre is still very much a work in progress. This difference in distances may imply the need for a further correction of ~ 0.2 mag when comparing LMC and bulge PNe luminosities.

Finally, we have also shown the completeness limits of the SAGE survey at 3.6 and 8.0 $\mu$m, below which source detection efficiencies are appreciably reduced. The values for these limits are taken from the SAGE delivery document at http://irsa.ipac.caltech.edu/data/SPITZER/SAGE/doc/SAGEDataDescription_Delivery2.pdf, and are important in defining the apparent distribution of LMC PNe.

Several differences may be noted between the distributions of bulge and LMC sources. It is clear for instance that the GBPNe lie predominantly in the range [3.6] > 9 mag and [3.6]-[8.0] < 3 – a regime which is below the 8 $\mu$m completeness limit of the LMC PNe. It is therefore apparent that if the [3.6] luminosity functions of the populations are in any way comparable, then we are failing to measure a very large fraction of the Magellanic PNe.

Note in this respect that differences of extinction are unlikely to affect this comparison between the sources. Even where there is a difference of $\Delta A_V$ = 10 mag between the extinctions of the bulge and LMC sources, for instance, the relative shifts of the populations would be reasonably modest (see the corresponding reddening vector in Fig. 1) We shall later provide evidence that differences in extinction may in fact be very much less (see Sect. 5). Similarly, although some of the scatter along the vertical axis may arise from differences in line-of-sight distance, this is unlikely to be greater than $\Delta[3.6]$ ~ 0.35 mag, and will be comparable for both of the populations.

Despite the fact that the LMC sample is far from complete, and the GBPNe may be magnitude limited because of corresponding radio detection limits, it is clear that the 3.6 $\mu$m luminosity functions of the bulge and Magellanic sources are very closely comparable, and peak close to [3.6] = 12.5 mag (Fig. 3). The somewhat slower fall-off of LMC sources to lower 3.6 $\mu$m luminosities (i.e. higher magnitudes [3.6]) may be attributable to the incompleteness of the sample having [3.6] > 12 mag – and the corresponding reduction that this causes in peak source fractions.

It may finally be noted that although the range in [3.6] magnitudes is of order ~5 mag for the GBPNe, there appears to be little correlation between the 3.6 $\mu$m magnitude on the one hand, and the [3.6]-[8.0] index on the other. The levels of dispersion, and mean value of [3.6] appear to be reasonably invariant and independent of the colour.

Somewhat similar results apply for the LMC sources as well, although our analysis is compromised by the relative incompleteness of the LMC sample.

The various tendencies noted above are also apparent when one compares the [8.0] and [3.6] magnitudes directly, as in Fig. 4. Although the sample of bulge nebulae is comparatively small (see the upper panel), that of the LMC nebulae is very much larger (lower panel).

It is apparent, from the LMC sample, that providing that [3.6] > 11 and [8.0] > 8, then there is very little variation in [3.6]-[8.0] colours as a function of either of these magnitudes. Furthermore, it is clear that the [3.6]-[8.0] index is very tightly constrained, and takes a maximum value close to ~3.8 mag. This presumably indicates the presence of an upper limit flux difference attributable to the 7.7 and 8.6 $\mu$m PAH band features.

The trend between the 4.5 and 3.6 $\mu$m magnitudes (black diamonds), by contrast, follows a linear relation about the free-free locus (indicated by the diagonal dashed line). It will be noted that there is a considerable scatter about this trend, with nebular points located along two closely parallel and overlapping lines. These two trends represent pure gaseous (upper trend), and dominant stellar emission (lower trend). The fact that the gaseous trend is ~0.3 mag higher than the bremsstrahlung relation presumably arises due to extra components of flux in the 4.5 photometric band, such as Br$\alpha$, [Ar VI] and [Mg IV], and some shift in values due to interstellar extinction.

The corresponding variation of [3.6]-[8.0] colours with 8.0 $\mu$m magnitudes is illustrated in Fig. 2. Here again, the limited completeness of the 8.0 $\mu$m SAGE sample greatly constrains the distribution of LMC sources, and very few nebulae possess [8.0] > 9.5 mag ($\equiv$ [8.0] > 13.5 mag at the distance of the LMC). This constraint upon sample sizes is more severe than was the case at 3.6 $\mu$m, and leads to a marked difference between the luminosity functions of the bulge and LMC sources (Fig. 3).

Finally, it is worth noting that the bulge sources show a systematic trend of [3.6]-[8.0] colour with 8.0 $\mu$m magnitude, a variation which is particularly evident where [3.6]-[8.0] > 3.0.

## 4. The Colour-Colour Diagrams

Unlike the case of the magnitude-colour diagrams, the colour-colour diagrams enable us to make direct comparisons between all four groupings of source considered here - the LMC PNe, the GBPNe, and the MASH and Acker et al. populations of the GDPNe. Even in this case, however, some care must be taken in interpreting the results, and trends within the colour plane may be influenced by the magnitude completeness limits described above.

We shall consider the distributions of sources within the [3.6]-[4.5]/[4.5]-[8.0] (Fig. 5), and [3.6]-[4.5]/[5.8]-[8.0] and [5.8]-[8.0]/[4.5]-[5.8] planes (Fig. 6) separately below.

**4.1 The [3.6]-[4.5]/[4.5]-[8.0] Colour Plane**

It is clear, from Fig. 5, that all of the sources occupy a similar regime of the [3.6]-[4.5]/[4.5]-[8.0] colour plane. There are however differences, and these help in defining the emission mechanisms responsible for these trends. In the first place, although it is clear that the ranges of [3.6]-[4.5] index are similar for all classes of nebulae, there is a tendency for LMC sources to be concentrated about [3.6]-[4.5] ~ 0.8 mag. This trend is even more apparent in Fig. 7, where we compare trends for all three categories of source. The nebulae in the upper three curves correspond to sources which have been detected in all four of the IRAC bands; that is, they correspond to exactly the same sample of PNe as is represented in Fig. 5. It is apparent, for these cases, that whilst the LMC sources peak within a range $\Delta$([3.6]-[4.5]) ~ 0.6 mag, the distributions for GDPNe and GBPNe are very much broader.

Some clue as to what might be happening may be determined when one plots the distribution of LMC sources having *no* 5.8 or 8.0 $\mu$m emission, as is the case for the lowermost curve of Fig. 7. In this case, the sources are again located within a narrow range of colours, but peak at indices close to [3.6]-[4.5] ~ 0.1 mag. There are several possible explanations for these trends.

In the first place, we note that nearly half of the Spitzer PNe appear to have no MIR dust emission at all (see e.g. Stanghellini et al. 2007); in which case 5.8 and 8.0 $\mu$m fluxes may fall below levels of Spitzer detectability. Such nebulae are also likely to have [3.6]-[4.5] indices which are quite distinct from those of their more dusty counterparts – and perhaps similar to what has been noted in our analysis above.

Models of radiatively accelerated AGB mass loss also suggest that low dust-producing progenitors may also have lower rates of mass-loss (see e.g. Winters et al. 2000; Sedlmayr & Dominik 1995); a situation which may result in lower levels of shorter-wave bremsstrahlung emission, and a dominance by the stellar continua.

Finally, we note that low [3.6]-[4.5] indices may also ensue where shell and progenitor masses are intrinsically smaller, and/or where the nebulae are highly evolved, shell surface brightness are small, and the nebulae fall, yet again, below levels of instrumental detectability (see our further discussion in Sect. 7).

By contrast, the source completeness limits for the bulge and disk sources are very much less severe, and the upper two curves of Fig. 7 contain both categories of source – nebulae in which shell emission is weak, and central star continua are relatively strong, and where the reverse is the case, and fluxes are dominated by shell components of emission alone. This leads to the prominent double-peaked profile for the bulge sources, and the comparatively broad distribution of sources noted for disk PNe.

Having said this, Hora et al. (2008) have noted that sources having large values of [4.5]-[8.0] also tend to have somewhat smaller indices [3.6]-[4.5] – a feature which is apparent in Fig. 5 where [4.5]-[8.0] > 3. They suggest that this may arise because of strong 3.3 $\mu$m PAH band emission, and/or the contribution of the Pf$\gamma$ transition at 3.74 $\mu$m, both of which are located within the 3.6 $\mu$m channel. Where this is the case, then it is apparent that it may contribute to sources having intermediate values of [3.6]-[4.5] (although the influence of Pf$\gamma$ is likely to be small, not least because increases in the strength of this particular transition will be more than offset by the contribution of Br$\alpha$ in the 4.5 $\mu$m band).

The distributions of sources with respect to the [4.5]-[8.0] colour axis, by contrast, appear to be much more similar – all classes of nebulae have similarly broad ranges of scatter. Here again, however, whilst the trends in source fraction (illustrated in Fig. 8) confirm that this is the case, there are also far from negligible differences. The LMC sources tend to be weighted towards larger values of the index, the bulge sources more inclined towards lower indices, and the GDPNe extend over the full range of indices, from the very small (~ 0.1) to the very large (~ 5).

These differences can, yet again, be partially attributed to the detection limits of the Spitzer SAGE survey. Given that we are including only those nebulae in which all four bands have been detected, it is apparent that LMC sources having lower 5.8 and/or 8.0 $\mu$m shell emission fall outside of the sample illustrated in Fig. 8. The excluded sources are likely to include a large fraction of nebulae in which central star continua are appreciable, and [4.5]-[8.0] colours are reduced.

## 4.2 The [3.6-[4.5]/[5.8]-[8.0] and [5.8]-[8.0]/[4.5]-[8.0] Colour Trends

The final two colour diagrams are illustrated in Fig. 6, and show the most radical differences in colour between the nebulae considered here. Considering firstly the [3.6]-[4.5]/[5.8]-[8.0] distribution, it is apparent that whilst the bulge and disk nebulae extend over a range -0.2 < [5.8]-[8.0] < 4, with a few outliers extending as high as [5.8]-[8.0] ~ 5, the LMC nebulae are strongly truncated to the range [5.8]-[8.0] < 2. This is even more clearly apparent in Fig. 8, where it is seen that the peak in LMC sources is narrow, and centred close to [5.8]-[8.0] ~ 1.75. There are at least three possible reasons for this disparity in colours. In the first place, a large fraction of LMC nebulae appear to possess no MIR dust emission at all. This is likely to introduce variations in colours similar to those noted above (see Sect. 6).

Most of the LMC nebulae appear to possess spectra dominated by carbon-rich dust emission features, however (Stanghellini et al. 2007). For these cases, it may be that 7.7 and 8.6 $\mu$m PAH features within the 8.0 $\mu$m band are very much weaker than the 6.2 $\mu$m feature within the 5.8 $\mu$m band. This may also be allied to weaker dust continuum emission with the 8.0 $\mu$m band.

Alternatively, it is possible that PAH emission within the 5.8 $\mu$m band is relatively stronger, and uniquely responsible for the reduction in [5.8]-[8.0] indices.

There is in fact some further evidence that this latter mechanism may be responsible for explaining these disparities. When one compares the [5.8]-[8.0] and [4.5]-[5.8] colours, as in the lower panel of Fig. 6, then it is apparent that whilst the [5.8]-[8.0] indices are much reduced within the LMC nebulae, the [4.5]-[5.8] indices are significantly higher. This suggests that it is the [5.8] magnitudes which may be appreciably truncated.

Finally, it is worth noting that the range of [4.5]-[8.0] indices is similar in all categories of source (see Figs. 5 & 8). Given that there is no PAH emission within the 4.5 μm photometric band, this suggest that levels of 7.7 and 8.6 μm PAH emission are at least comparable, if not necessarily the same.

It is therefore possible that the strength of the 6.2 μm PAH and plateau features are relatively larger in LMC nebulae as compared to their strengths in the Galactic bulge and disk PNe.

We shall discuss these trends further in Sect. 7, where it will be pointed out that a variety of mechanisms may be responsible for modifying the PAH emission band strengths.

A final interesting aspect of the distribution in Fig. 6 relates to the Galactic disk components themselves. It will be noted that the Acker et al. (1992) disk sources, represented by the black diamonds, have a distribution which is not very much different from that of their MASH counterparts (represented by the squares). This result differs from that noted in previous studies of these nebulae (Cohen et al. 2007a), where it was determined that the colour range of MASH PNe was very much smaller than those of the Acker et al. sources.

**5. Near-infrared Colour distribution**

The distribution of 2MASS indices for bulge and LMC PNe is illustrated in Fig. 9, wherein we also indicate the trends observed for GDPNe (Ramos-Larios & Phillips 2005). The indices include all of the sources, irrespective of whether they were detected in longer wavelength Spitzer photometry or not, and are located within a couple of arcsecs of the specified PNe positions.

It can be seen that whilst a reasonable fraction of the colours appear to correspond to heavily reddened PNe bremsstrahlung emitting shells (most of those having H-$K_S$ > 0.65 mag), the larger part of the LMC detections appear to concentrate close to J-H ~ 0.5 mag, H-$K_S$ ~ 0.15 mag, or along a diagonal line to the upper right-hand side of this grouping. If our identifications are correct, it would therefore seem likely that most of these sources have fluxes dominated by the central star continua, and are experiencing varying levels of interstellar reddening. Where this is the case, and one employs the extinction coefficients of Gordon et al. (2003), then this would give LMC extinctions of $\approx$7 mag or so.

The bulge nebulae, on the other hand, tend to be congregated close to J-H ~ 1.25 mag, H-$K_S$ ~ 0.54 mag – a situation that can again be interpreted in terms of stellar dominated continua, and high levels of extinction. Where this is the case, and assuming that reddening can be approximated by the prescription of Cardelli et al. (1989) for $R_V$ = $A_V/E_{B-V}$ = 3.1, then this implies that extinction for the main body of the sources is of order $A_V$ = 11.6 mag. The differential extinction between these groups of nebulae, the LMC sources on the one hand and the GBPNe on the other, would then appear to be of order $\Delta A_V$ ~ 4.6 mag. However, it should be noted that there is evidence that values of $R_V$ towards the Galactic centre may be much less than supposed above, with Walton et al. (1993) finding $R_V$ = 2.3, and Ruffle et al. (2004) determining $R_V$ = 2.0. If this latter value is adopted, then mean values of $A_V$ would climb to ~15 mag or so.

All of these estimates of extinction seem to be somewhat on the large side, however. For instance, the bulge sources of Gutenkunst et al. appear to have typical extinctions of close to C ~ 2, which would imply $A_V$ ~ 4.25 mag for $R_V$ = 3.1. Similarly, we note that LMC planetaries seem to have typical extinctions <$E_{B-V}$> $\cong$ 0.13 mag (Villaver et al. 2003, 2004), which might imply values $A_V$ $\cong$ 0.44 where one uses $R_V$ = 3.41 (Gordon et al. 2003). So on this basis, and with these lower extinctions, it is clear that the difference in extinction would be of order $A_V$(GBPNe)-$A_V$(LMC) ~3.8 mag, and somewhat larger than this if the lower values of $R_V$ for the Galactic bulge are employed.

So it seems that whatever way one spins the coin, the mean difference in extinction between bulge and LMC PNe is likely to be ~4 mag or so. However, the difference in the absolute values of these extinctions is certainly troubling, and brings into question whether the supposedly stellar dominated sources might not contain appreciable levels of plasma emission, or whether some of these apparent identifications refer to differing sources altogether; perhaps later G-M stars in the line-of-sight. It follows that the present 2MASS results, and the corresponding NIR photometry of Hora et al. (2008) should be treated with a certain degree of caution.

## 6. Discussion

It is clear, from Sects. 4 & 6, that we are observing a broad range of shell emission mechanisms within the MIR; processes which result in bremsstrahlung emission, central star continuum emission (both of the latter particularly strong at 3.6 and 4.5 $\mu$m), permitted and forbidden line emission, and (most probably) various shocked and/or fluorescently excited transitions of $H_2$. The dominant components at longer wavelengths, however, are likely to be the PAH band features, gaseous thermal continuum emission, and underlying smooth components of continuum arising from amorphous carbon or silicate grains. Given that many of the latter particles are located within PDRs, and that the PDRs may extend a considerable distance from the central ionised zones, then this may explain the broad increases in source size observed at longer IRAC wavelengths, detected both in the present nebulae (Phillips & Ramos-Larios 2009, in preparation) and in other PNe (see e.g. Phillips & Ramos-Larios 2008a, b).

Such mechanisms also go a long way to explaining the large ranges of colour index noted in Figs. 5-6. Generally speaking, whilst dust continuum and PAH emission will tend to shift PNe towards larger values of [4.5]-[8.0], $H_2$ and ionized gas emission will tend to force sources in a reverse direction. Similarly, whilst PAH and ionized gas emission tends to reduce indices [3.6]-[4.5], dust continuum emission would tend to lead to a reverse effect.

Given that various of these mechanisms are compositionally dependent, then it might be anticipated that the distributions of colours might also vary between the groups of sources considered here. Thus for instance, it is well established that PN abundances in the Magellanic Clouds follow the trends detected for a variety of other sources, including stars and HII regions, and imply abundances which are significantly less than those of their Galactic counterparts. Such differences in abundance are also likely to be reflected in the peculiarities of their MIR spectra, noted for instance in the Spitzer analysis of Stanghellini et al. (2007). It appears from this that nearly half of their sources show no MIR dust emission at all, and that MIR indices are likely to be dominated by gas thermal continua and line contributions. Where dust band emission is detected, then it appears that most of the sources contain SiC or PAH emission components; the incidence of silicate features is relatively small.

Similarly, whilst the metallicities of GBPNe are somewhat larger than those for GDPNe, it appears that carbon abundances may be somewhat less (Wang & Liu 2007; Casassus et al. 2001). This may contribute to the larger incidence of silicate features in the MIR spectra of these sources – indeed, they are present in all of the GBPNe which have been analysed so far (Gutenkunst et al. 2008; Perea-Calderón et al. 2009).

Having said this, however, it is apparent that the bulge sources contain a larger proportion of nebulae in which carbon-rich features (e.g. the SiC band at 10.5-12.7 mm; see e.g. Forrest et al. 1975; Anderson et al. 1999; Casassus et al. 2001) and oxygen-rich features (such as the crystalline silicate features at 26-37 $\mu$m; see e.g. Molster et al. 2002) are observed within the self-same shells (Gutenkunst et al. 2008; Perea-Calderón et al. 2009) – a trend which appears to be much more common than is the case for GDPNe. This may arise because the incidence of binary induced morphologies is larger in the bulge (viz. Zijlstra 2007), and as a result of the tendency for many bipolar sources to have inner C-rich bipolar outflows, surrounded by outer O-rich tori (see e.g. the discussion of Gutenkunst et al. 2008). The crystalline silicate features may arise where AGB winds impact these tori and anneal the dust (Edgar et al. 2008). It has also been suggested that late thermal pulses at the end of the AGB phase may lead to changes in C/O ratios within the flows (Waters et al. 1998); that evaporation of Oort-belt comets may release crystalline silicates (Cohen et al. 1999); and that various other processes involving binary systems, variations in hot-bottom burning, and brown dwarfs or planets may also explain these trends (see the slightly more detailed discussions of Gutenkunst et al. (2008) and Perea-Calderón et al. (2009), and references therein). It is therefore clear that there is no shortage of explanations, but still as yet little understanding of these trends.

So, there is considerable evidence for compositional differences between GDPNe, GBPNe and LMC PNe, and the question arises as to whether such variations are also reflected in the colour-colour diagrams described in Sect. 4.

We have already pointed out that there appear to be marked differences in distribution with respect to the [3.6]-[4.5] and [4.5]-[8.0] indices (see Figs. 5-8). The LMC sources tend to be more narrowly peaked, and located towards higher values of these indices, for instance. It was pointed out however that such trends may occur where one is excluding shells in which MIR dust emission is reduced, and 5.8 $\mu$m and/or 8.0 $\mu$m intensities are small; a result which applies to nearly half of the LMC nebulae investigated by Stanghellini et al. (2007) This results in the elimination of sources in which stellar continua may be important, and/or possess 3.6 and 4.5 $\mu$m fluxes which are affected by a differing balance of emission mechanisms (e.g. reduced bremsstrahlung and/or 3.2 $\mu$m PAH emission), leading to [3.6]-[4.5] and [4.5]-[8.0] indices which are correspondingly reduced.

A similar analysis for more evolved sources suggests that comparable trends would apply for these sources as well. As nebulae evolve towards their turn-over points in the HR plane, where hydrogen burning ceases, then central star temperatures increase (e.g. Vassiliadis & Wood 1994)) and absolute magnitudes M become larger (i.e. MIR and visual fluxes decrease; e.g. Schonberner 1981), even though bolometric luminosities remain more-or-less the same. Beyond this stage, however, the central star luminosities and temperatures go into a sharp decline, stellar absolute magnitudes remain more-or-less invariant, and shell surface brightnesses decline as $\propto R^{-5}$ (where R is the radius of the nebula). The surface brightnesses of the shells are therefore expected to decline very rapidly indeed, and will eventually fall below levels of instrumental detectability. The stellar MIR fluxes, by contrast, would remain more-or-less invariant, and eventually dominate fluxes in the shorter wave IRAC passbands.

It therefore follows that the exclusion of intrinsically fainter (or lower surface brightness) nebulae removes sources which would normally occupy lower ranges of the [3.6]-[4.5] and [4.5]-[8.0] colour regimes.

Having stated all of this, however, and taken note of other observationally driven biases as well, it is nevertheless apparent that there are intrinsic similarities and differences between the various nebular groupings. On the one hand, it is clear that despite the partial sampling of LMC nebulae, the 3.6 $\mu$m luminosity function for these sources is similar to that of the GBPNe. Similarly, it is noted that the LMC PNe, GBPNe and GDPNe extend over similar ranges of index [4.5]-[8.0] – although there appear to be proportionately more GDPNe at the very highest values of this index (> 4).

Perhaps the sharpest disparities, however, are those noted in the [5.8]-[8.0] and [4.5]-[5.8] colours. There are at least two possibilities for explaining these results. In the first place, and as noted above, nearly half of the LMC spectra of Stanghellini et al. (2007) appear to contain no MIR dust emission at all – a complete contrast to what has so far been observed for Galactic bulge and disk PNe. So for these sources, at least, it is clear that emission would be dominated by bremsstrahlung continuum and forbidden line emission, including the strong transitions of [ArII] 6.99 $\mu$m, [ArV] 7.90 $\mu$m and [ArIII] 8.99 $\mu$m. Permitted transitions of HI, and various transitions due to $H_2$ may also contribute to the mix. The net result would be for limits upon the [5.8]-[8.0] and [4.5]-[5.8] indices which are consistent with those noted in Fig. 6. However, this situation is likely to apply to less than half of the LMC sources considered here, whilst many of them are likely to have 5.8 and 8 $\mu$m fluxes which exclude from the colour-colour analysis described above (i.e. causes them to fall below the Spitzer SAGE detection limits). The question then arises as to how one might explain the carbon dominated spectra of most of the other PNe – that is, those spectra in which PAH features dominate our present IRAC fluxes.

To understand what might be happening in the latter case, we note first that emission at 4.5 $\mu$m is likely to be influenced by various atomic and ionic transitions (such as Br$\alpha$, the forbidden lines [Ar VI] and [Mg IV] close to 4.53 $\mu$m, and the shock or fluorescently excited v=0-0 S(8) and S(9) lines of $H_2$ at $\lambda\lambda$4.69 and 5.05 $\mu$m), as well as by free-free and bound-free plasma emission. There are however no PAH band features, such as are observed in the other IRAC passbands. It therefore follows that the [5.8]-[8.0] and [4.5]-[5.8] indices are likely telling us something about the relative strengths of the 6.2 PAH feature in the 5.8 $\mu$m band, and the 7.7 and 8.6 $\mu$m features in the 8.0 $\mu$m band – not to mention the related, and much broader plateau components which are associated with these contributions. It was argued, in Sect. 4.2, that the fact that the LMC sources tend to have highly restricted (and low) values of [5.8]-[8.0], larger values of [4.5]-[5.8], but rather similar indices [3.6]-[8.0], may be indicative of enhanced 6.2 µm PAH emission – that the LMC sources have larger 6.2µm/(7.7µm+8.6µm) PAH ratios than is the case for the GDPNe and GBPNe

This, should it be the case, may arise from differences in the natures and dimensions of the PAH particles, and depend upon whether they contain smaller aggregates of C atoms such as coronene, or represent larger structures such as dicoronene (see e.g. Léger & d'Hendecourt 1988). It also depends upon whether the PAH atoms are neutral or ionised.

Where ionisation is important, for instance, then this will tend to enhance C-C stretching vibrations, such as are principally responsible for the 6.2 µm PAH band feature (see e.g. Peeters et al. 2002, and the related laboratory work of Allamandola et al. 1999). The transitions responsible for the 7.7 and 8.6 µm features, by contrast, correspond to C-H in-plane bending modes (in the case of the 8.6 µm feature), and combined C-H in-plane bending and C-C stretching modes (for the remaining transitions). Ionisation of the particles may therefore shift PAH strengths in the direction noted for our LMC nebulae.

It has also been noted that small PAH grains and clusters possess differing levels of emission for the 6.2 µm feature on the one hand, and the 7.7 and 8.6 µm features on the other, with ratios I(7.7µm+8.6µm)/I(6.2µm) tending to be larger in the case of larger PAH clusters having 100-1000 C atoms (see e.g. Rapacioli, Joblin & Boissel 2005; Berné et al. 2008).

It would therefore appear that a multiplicity of mechanisms might cause variations in relative PAH emission band strengths, and explain the differences in the MIR colour planes noted in this work. Broadly speaking, where the properties of grain populations vary, and/or conditions of ionisation are modified, then one might expect corresponding changes in relative PAH band strengths.

Such variations in grain properties might be expected to occur where there are changes in source metallicity, such as appear to arise between the LMC and Galactic PN outflows (see our comments in Sect. 1), whilst one might also expect to observe differences in underlying levels of grain continuum emission; in central star properties (see e.g. the evolutionary curves of Vassiliadis & Wood (1994) for differing progenitor metallicities); in line emission characteristics, and, conceivably, in the masses of the nebular shells.

Thus for instance, given that dust/gas mass ratios appear to be surprisingly similar for all metallicities of outflow (see e.g. Phillips 2007), then it is arguable that Magellanic sources would show a higher level of elemental depletion – that is, an increase in the fractional masses of heavier elements which are contained within the dust, and a corresponding reduction in gas phase abundances for the lower metallicity sources (Phillips 2007). This would likely lead to corresponding changes in the permitted and forbidden line spectra.

Similarly, differences in metallicity would plausibly influence grain formation properties within the flows (see e.g. the discussion by Ferrarotti & Gail (2006) of dust formation in C- and O-rich environments, and Sedlmayr & Dominik (1995) for the physics and chemistry involved radiatively driven outflows), and this may account for the lower fractional masses of PAH type grains in lower metallicity environments (see e.g. Draine et al. 2007).

Further testing of such hypotheses is possible through MIR spectroscopy of the sources, and it would be of interest to obtain a larger sample of such observations than is currently available.

Finally, and as mentioned above, it is fascinating to note that [4.5]-[8.0] indices for LMC and Galactic nebulae are reasonably similar. This is perhaps somewhat unexpected given our discussion above – although it is possible that a combination of lower shell masses and PAH fractional masses may lead to compensating reductions in 4.5 and 8.0 $\mu$m emission, and comparable values for [4.5]-[8.0].

## 7. Conclusions

We have undertaken a comparative analysis of the magnitudes and colours of GBPNe, GDPNe and LMC PNe based upon data products deriving from the GLIMPSE II and SAGE Galactic Legacy Programs. This has permitted us to determine that the 3.6 $\mu$m luminosity function for LMC PNe is similar to that of their Galactic bulge counterparts – and within the uncertainties engendered by incompleteness of the LMC sample. The colour-colour distributions of the sources are also for the most part similar, although with differences (in the case of [3.6]-[4.5]) arising from incompleteness of the LMC sample.

The only exceptions to this trend appear to be in the [4.6]-[5.8] and [5.8]-[8.0] indices. The LMC nebulae tend to be confined to higher values of the [4.6]-[5.8] index, for instance – much larger than is the case for the bulge and disk PNe. By contrast, the [5.8]-[8.0] index tends to be significantly lower. This difference between the LMC and Galactic sources may arise due to enhanced PAH band emission within the 5.8 μm photometric channel, suggesting that the nature of PAH excitation, or of the PAH grains themselves, may differ between these differing categories of source. Alternatively, such a situation may reflect the fact that a large proportion of LMC nebulae appear to contain very little MIR dust emission at all, leading to a dominance by permitted and forbidden line fluxes, stellar continua, and bremsstrahlung emission.

Previous analyses of MASH sources appeared to indicate that many of them were located within a highly restricted regime of the [3.6]-[4.5]/[5.8]-[8.0] colour plane; a regime which was much more constrained than that of less evolved GDPNe, such as the sources listed in the catalogue of Acker et al. (1992).

Our present results do not confirm this trend, but rather indicate that the Acker et al. (1992) and MASH sources have similar colours.

**Acknowledgements**

We would like to thank an anonymous referee for several very useful insights. This referee also suggested removing a large section of mapping data relating to the present sources; data which will now be published in a further article concerning GLIMPSE II and 3D PNe. This work is based, in part, on observations made with the Spitzer Space Telescope, which is operated by the Jet Propulsion Laboratory, California Institute of Technology under a contract with NASA. Support for this work was provided by an award issued by JPL/Caltech. It also makes use of data products from the Two Micron All Sky Survey, which is a joint project of the University of Massachusetts and the Infrared

Processing and Analysis Center/California Institute of Technology, funded by the National Aeronautics and Space Administration and the National Science Foundation. GRL acknowledges support from CONACyT (Mexico) grant 93172.

<a>
<p><b>Planetary Nebulae II: From Origins to Microstructures. Astron. Soc. Pac., San Francisco, p. 397</b></p></a>

Table 1

MIR Photometry of Acker and MASH Planetary Nebulae

| G.C. | CAT. | NAME | [3.6] mag | $\sigma_{3.6}$ mag | [4.5] mag | $\sigma_{4.5}$ mag | [5.8] mag | $\sigma_{5.8}$ mag | [8.0] mag | $\sigma_{8.0}$ mag |
|---|---|---|---|---|---|---|---|---|---|---|
| G000.0-01.3 | MASHI | PPA1751-2933 | 12.92 | 0.05 | 12.08 | 0.04 | 10.73 | 0.10 | 7.31 | 0.01 |
| G000.1-01.7 | MASHI | PHR1752-2941 | 11.71 | 0.08 | 11.37 | 0.02 | 11.52 | 0.07 | 7.92 | 0.01 |
| G000.1-01.1 | ACK | M 3-43 | 10.43 | 0.30 | 9.53 | 0.09 | 8.29 | 0.07 | 5.85 | 0.06 |
| G000.1+01.9 | MASHI | PHR1738-2748 | 11.51 | 0.03 | 11.68 | 0.05 | 11.69 | 0.06 | 11.20 | 0.23 |
| G000.2-01.9 | ACK | M 2-19 | 10.65 | 0.14 | 10.83 | 0.13 | 9.94 | 0.09 | 9.00 | 0.01 |
| G000.4-01.9 | ACK | M 2-20 | 10.16 | 0.11 | 9.80 | 0.12 | 9.18 | 0.06 | 7.13 | 0.06 |
| G000.5-01.6 | ACK | Al 2-Q | 12.54 | 0.05 | 11.82 | 0.03 | 12.92 | 0.06 | 10.12 | 0.01 |
| G000.6-01.3 | ACK | Bl 3-15 | 10.20 | 0.02 | 10.55 | 0.02 | 9.59 | 0.01 | 8.48 | 0.02 |
| G000.8-01.5 | ACK | Bl O | 10.95 | 0.02 | 11.06 | 0.01 | 9.47 | 0.01 | 8.14 | 0.02 |
| G000.9-02.0 | ACK | Bl 3-13 | 12.67 | 0.14 | 11.36 | 0.11 | 11.86 | 0.05 | 8.71 | 0.09 |
| G001.0-01.9 | MASHI | PHR1755-2904 | 13.58 | 0.38 | 11.36 | 0.01 | 11.76 | 0.04 | 7.39 | 0.02 |
| G001.0+01.9 | ACK | K 1- 4 | 9.67 | 0.04 | 9.65 | 0.05 | 9.69 | 0.06 | 9.59 | 0.03 |
| G001.1-01.6 | ACK | Sa 3- 92 | 11.86 | 0.01 | 12.30 | 0.03 | 12.56 | 0.04 | 10.88 | 0.01 |
| G001.1-01.2 | MASHI | PPA1753-2836 | 7.89 | 0.02 | 8.15 | 0.02 | 7.89 | 0.02 | 7.07 | 0.02 |
| G001.3-01.2 | ACK | Bl M | 11.12 | 0.04 | 10.91 | 0.01 | 10.41 | 0.02 | 8.77 | 0.02 |
| G001.7-01.6 | ACK | H 2-31 | 11.36 | 0.11 | 10.63 | 0.07 | 9.52 | 0.06 | 7.38 | 0.06 |
| G002.1-00.9 | MASHI | PHR1754-2736 | 10.16 | 0.02 | 9.46 | 0.02 | 8.54 | 0.02 | 5.59 | 0.02 |
| G002.2-01.2 | MASHI | PPA1755-2739 | 13.10 | 0.08 | 11.25 | 0.01 | 10.82 | 0.05 | 7.11 | 0.01 |
| G002.2+00.5 | ACK | Te 2337 | 10.48 | 0.02 | 9.42 | 0.02 | 8.02 | 0.02 | 4.99 | 0.02 |
| G003.5+01.3 | MASHII | MPA1748-2511 | 12.25 | 0.01 | 10.87 | 0.01 | 11.27 | 0.03 | 9.33 | 0.02 |
| G003.6-01.3 | MASHI | PHR1759-2630 | 11.20 | 0.01 | 10.14 | 0.03 | 10.05 | 0.01 | 7.25 | 0.02 |
| G004.0-00.4 | MASHI | PHR1756-2538 | --- | --- | 9.19 | 0.02 | 8.88 | 0.04 | 7.34 | 0.01 |
| G004.3-01.4 | MASHI | PPA1801-2553 | 12.09 | 0.02 | 11.11 | 0.01 | 11.09 | 0.08 | 7.40 | 0.01 |
| G004.8-01.1 | MASHI | PHR1801-2522 | 10.47 | 0.02 | 9.39 | 0.01 | 9.20 | 0.05 | 5.85 | 0.01 |
| G006.1+00.8 | MASHI | PPA1756-2311 | 12.71 | 0.08 | 11.94 | 0.04 | 12.24 | 0.03 | 7.20 | 0.02 |
| G006.2+01.0 | ACK | HaTr 8 | 13.20 | 0.10 | 12.54 | 0.16 | 12.15 | 0.10 | 10.54 | 0.04 |
| G008.6+01.0 | MASHI | PHR1801-2057 | --- | --- | 12.38 | 0.04 | --- | --- | 7.04 | 0.02 |
| G352.6+00.1 | ACK | H 1-12 | 8.14 | 0.02 | 7.59 | 0.02 | 7.82 | 0.02 | 6.42 | 0.02 |
| G352.8-00.5 | MASHII | MPA1729-3513 | 9.47 | 0.01 | 8.39 | 0.02 | 7.45 | 0.02 | 5.99 | 0.02 |
| G352.8-00.2 | ACK | H 1-13 | 8.16 | 0.02 | 7.57 | 0.02 | 7.88 | 0.02 | 6.13 | 0.01 |
| G353.9+00.0 | MASHI | PPA1730-3400 | 9.55 | 0.01 | 8.98 | 0.02 | 8.48 | 0.01 | 7.53 | 0.01 |
| G355.6+01.4 | MASHI | PPA1729-3152 | 11.46 | 0.01 | 10.47 | 0.02 | 9.27 | 0.02 | 7.23 | 0.02 |
| G356.0-01.4 | MASHI | PPA1741-3302 | 13.01 | 0.06 | 11.51 | 0.03 | 11.68 | 0.11 | 8.30 | 0.01 |
| G356.5+01.5 | ACK | Th 3-55 | 12.12 | 0.02 | 11.53 | 0.02 | 11.63 | 0.03 | 9.99 | 0.02 |
| G356.9+00.9 | MASHI | PPA1734-3102 | 10.33 | 0.02 | 10.10 | 0.01 | 8.36 | 0.01 | 6.17 | 0.02 |
| G357.2+01.4 | ACK | Al 2-H | 13.78 | 0.16 | 12.63 | 0.06 | 12.94 | 0.12 | 11.34 | 0.11 |
| G357.5+01.3 | MASHI | PPA1734-3015 | 11.61 | 0.02 | 11.27 | 0.03 | 9.04 | 0.01 | 7.15 | 0.01 |
| G358.2-01.1 | ACK | Bl D | 10.72 | 0.02 | 9.86 | 0.02 | 9.72 | 0.02 | 8.05 | 0.02 |
| G358.3+01.2 | ACK | Bl B | 10.56 | 0.01 | 10.59 | 0.02 | 9.72 | 0.01 | 8.21 | 0.01 |
| G358.8-00.0 | ACK | Te 2022 | 5.50 | 0.02 | 5.37 | 0.02 | 2.37 | 0.02 | 0.55 | 0.02 |
| G359.1-01.7 | ACK | M 1-29 | 9.79 | 0.02 | 8.96 | 0.01 | 9.03 | 0.02 | 7.34 | 0.02 |
| G359.2+01.2 | ACK | 19W32 | 8.83 | 0.01 | 8.84 | 0.02 | 6.93 | 0.02 | 6.19 | 0.02 |
| G359.3+01.4 | ACK | Th 3-35 | 10.92 | 0.10 | 9.62 | 0.07 | 8.95 | 0.05 | 6.55 | 0.04 |
| G359.3-00.9 | ACK | Hb 5 | 7.30 | 0.02 | 6.30 | 0.02 | 5.02 | 0.02 | 2.82 | 0.02 |
| G359.3-01.8 | ACK | M 3-44 | 10.39 | 0.23 | 9.88 | 0.13 | 8.23 | 0.07 | 6.57 | 0.14 |
| G359.7-01.8 | ACK | M 3-45 | 11.87 | 0.02 | 11.03 | 0.02 | 11.24 | 0.06 | 10.83 | 0.04 |

Table 2

MIR Photometry of the Galactic Bulge Planetary Nebulae
of Van de Steene & Jacoby (2001)

| G.C. | JaSt | [3.6] mag | $\sigma_{3.6}$ mag | [4.5] mag | $\sigma_{4.5}$ mag | [5.8] mag | $\sigma_{5.8}$ mag | [8.0] mag | $\sigma_{8.0}$ mag |
|---|---|---|---|---|---|---|---|---|---|
| G000.01-1.80 | 83 | --- | --- | 11.09 | 0.20 | --- | --- | 11.16 | 0.14 |
| G000.05+1.29 | 27 | 11.20 | 0.05 | 11.26 | 0.09 | 10.99 | 0.07 | 10.18 | 0.14 |
| G000.08-0.93 | 67 | 11.87 | 0.19 | 10.77 | 0.24 | 10.67 | 0.24 | 8.92 | 0.05 |
| G000.10-1.91 | 93 | 13.71 | 0.09 | 13.30 | 0.11 | 13.58 | 0.22 | 10.68 | 0.03 |
| G000.15-1.73 | 85 | 11.61 | 0.00 | 11.79 | 0.11 | 12.21 | 0.05 | 10.88 | 0.02 |
| G000.17-1.21 | 75 | 11.61 | 0.10 | 10.72 | 0.11 | 10.37 | 0.08 | 8.49 | 0.05 |
| G000.18-1.04 | 69 | 11.31 | 0.11 | 10.67 | 0.07 | 10.37 | 0.08 | 9.64 | 0.06 |
| G000.20-1.47 | 79 | 5.08 | 0.09 | 4.82 | 0.09 | 3.91 | 0.03 | 3.40 | 0.05 |
| G000.28+1.71 | 19 | 13.07 | 0.11 | 12.70 | 0.27 | 12.34 | 0.04 | 11.48 | 0.04 |
| G000.33-1.64 | 86 | 12.59 | 0.06 | 11.84 | 0.07 | 12.63 | 0.21 | 11.00 | 0.01 |
| G000.34+1.56 | 23 | 11.86 | 0.09 | 10.49 | 0.07 | 8.77 | 0.04 | 6.08 | 0.04 |
| G000.35+1.70 | 21 | 10.28 | 0.05 | 10.38 | 0.06 | 10.44 | 0.08 | 10.27 | 0.06 |
| G000.39+0.63 | 49 | 12.51 | 0.04 | 12.38 | 0.04 | 12.27 | 0.03 | 10.83 | 0.02 |
| G000.49+1.12 | 36 | 11.77 | 0.13 | 10.80 | 0.10 | 10.43 | 0.10 | 9.83 | 0.11 |
| G000.54+1.91 | 17 | 12.90 | 0.08 | 12.06 | 0.03 | 12.86 | 0.12 | 10.12 | 0.01 |
| G000.59-1.76 | 96 | 9.89 | 0.06 | 10.01 | 0.08 | 10.01 | 0.06 | 9.93 | 0.05 |
| G000.62-1.04 | 77 | 10.17 | 0.12 | 9.29 | 0.05 | 8.92 | 0.04 | 6.49 | 0.04 |
| G000.74-0.86 | 74 | 12.08 | 0.15 | 11.10 | 0.13 | 11.05 | 0.15 | 9.52 | 0.12 |
| G000.78-0.74 | 70 | 11.82 | 0.05 | 11.10 | 0.01 | 12.20 | 0.11 | 10.29 | 0.00 |
| G000.82+1.30 | 38 | 12.28 | 0.15 | 12.24 | 0.13 | 11.53 | 0.23 | 10.30 | 0.06 |
| G000.86-0.69 | 71 | --- | --- | 10.14 | 0.12 | --- | --- | 8.06 | 0.08 |
| G000.90+1.13 | 44 | 12.58 | 0.10 | 12.43 | 0.15 | 12.39 | 0.24 | 9.85 | 0.14 |
| G000.94-0.91 | 78 | 12.04 | 0.01 | 11.59 | 0.01 | 12.41 | 0.04 | 11.23 | 0.01 |
| G000.95-0.78 | 76 | 12.82 | 0.11 | 11.17 | 0.09 | 11.12 | 0.08 | 10.27 | 0.08 |
| G001.02+1.35 | 41 | 12.19 | 0.12 | 11.51 | 0.16 | 10.86 | 0.11 | 9.41 | 0.09 |
| G001.13+0.80 | 54 | 8.89 | 0.00 | 8.41 | 0.00 | 5.95 | 0.00 | 3.89 | 0.00 |
| G001.17+1.68 | 34 | 13.62 | 0.14 | 13.21 | 0.17 | 11.97 | 0.29 | 9.22 | 0.01 |
| G001.23+0.71 | 56 | 11.41 | 0.12 | 11.09 | 0.17 | 10.89 | 0.18 | 9.85 | 0.26 |
| G001.23+1.33 | 45 | 13.17 | 0.10 | 13.11 | 0.12 | --- | --- | --- | --- |
| G001.28-1.26 | 95 | 13.17 | 0.10 | 12.13 | 0.11 | 13.23 | 0.07 | 10.35 | 0.01 |
| G001.38-1.08 | 89 | 12.29 | 0.07 | 11.97 | 0.02 | 12.77 | 0.03 | 11.39 | 0.06 |
| G001.58+1.51 | 46 | 11.83 | 0.09 | 10.59 | 0.06 | 10.08 | 0.05 | 7.76 | 0.04 |
| G001.60-1.00 | 90 | 11.00 | 0.11 | 11.02 | 0.12 | 10.40 | 0.10 | 9.98 | 0.19 |
| G001.60+1.58 | 42 | 12.43 | 0.02 | 11.71 | 0.24 | 11.32 | 0.13 | 8.26 | 0.06 |
| G001.71+1.30 | 52 | 11.11 | 0.05 | 10.37 | 0.06 | 10.18 | 0.06 | 7.77 | 0.03 |
| G001.86-0.53 | 81 | 11.05 | 0.06 | 9.95 | 0.05 | 9.05 | 0.06 | 7.15 | 0.03 |
| G002.03-1.38 | 98 | 10.91 | 0.03 | 9.97 | 0.05 | 8.47 | 0.02 | 6.43 | 0.02 |
| G358.02+1.56 | 1 | 13.31 | 0.17 | 13.07 | 0.06 | 13.60 | 0.16 | 12.35 | 0.01 |
| G358.17+0.55 | 11 | 12.84 | 0.05 | 12.78 | 0.05 | 12.68 | 0.07 | 10.63 | 0.16 |
| G358.31+0.27 | 24 | 11.71 | 0.01 | 11.55 | 0.03 | 11.70 | 0.03 | 9.86 | 0.01 |
| G358.40+1.73 | 2 | 12.50 | 0.15 | 12.30 | 0.16 | 10.87 | 0.09 | 10.30 | 0.01 |
| G358.44+1.66 | 3 | 12.70 | 0.12 | 12.06 | 0.25 | 11.22 | 0.14 | 11.02 | 0.04 |
| G358.51-1.74 | 64 | 10.34 | 0.04 | 9.38 | 0.05 | 8.63 | 0.03 | 7.01 | 0.03 |
| G358.60+1.70 | 4 | 14.01 | 0.22 | 12.88 | 0.06 | 13.97 | 0.19 | 11.18 | 0.05 |
| G358.63+0.75 | 16 | 11.88 | 0.16 | 10.65 | 0.16 | 10.92 | 0.16 | 8.78 | 0.21 |

| G.C. | JaSt | [3.6] mag | $\sigma_{3.6}$ mag | [4.5] mag | $\sigma_{4.5}$ mag | [5.8] mag | $\sigma_{5.8}$ mag | [8.0] mag | $\sigma_{8.0}$ mag |
|---|---|---|---|---|---|---|---|---|---|
| G358.69-1.11 | 58 | 12.49 | 0.07 | 11.87 | 0.07 | 11.56 | 0.05 | 9.69 | 0.06 |
| G358.71+0.49 | 26 | 12.90 | 0.10 | 12.21 | 0.09 | 11.53 | 0.06 | 9.95 | 0.11 |
| G358.83+1.78 | 5 | 12.97 | 0.07 | 12.02 | 0.02 | 11.94 | 0.08 | 11.72 | 0.02 |
| G358.99-1.55 | 65 | 10.86 | 0.05 | 9.85 | 0.05 | 9.16 | 0.04 | 5.58 | 0.04 |
| G359.02+1.16 | 9 | 12.66 | 0.07 | 11.83 | 0.02 | 11.90 | 0.06 | 10.03 | 0.07 |
| G359.23+1.36 | 8 | 13.79 | 0.25 | 12.72 | 0.07 | 13.68 | 0.07 | 10.79 | 0.01 |
| G359.29+1.41 | 7 | 11.89 | 0.09 | 11.72 | 0.09 | 11.68 | 0.14 | 11.81 | 0.14 |
| G359.52-1.36 | 68 | 12.12 | 0.07 | 11.10 | 0.07 | 10.46 | 0.08 | 7.39 | 0.04 |
| G359.52-1.24 | 66 | 10.92 | 0.19 | 10.20 | 0.13 | 9.56 | 0.07 | 8.62 | 0.06 |
| G359.57+0.80 | 31 | 12.28 | 0.02 | 11.93 | 0.04 | 12.42 | 0.07 | 10.08 | 0.07 |
| G359.67-0.77 | 60 | 9.80 | 0.05 | 8.84 | 0.09 | 7.42 | 0.03 | 5.35 | 0.04 |
| G359.76-1.45 | 73 | 11.27 | 0.02 | 11.21 | 0.18 | 10.76 | 0.10 | 8.53 | 0.05 |

Table 3

2MASS and Spitzer Photometry for the LMC Planetary Nebulae of Reid & Parker (2006)

| # | l deg | b deg | NAME | STAT | J mag | σ_J mag | H mag | σ_H mag | K_S mag | σ_KS mag | [3.6] mag | σ_3.6 mag | [4.5] mag | σ_4.5 mag | [5.8] mag | σ_5.8 mag | [8.0] mag | σ_8.0 mag | r arcs |
|---|---|---|---|---|---|---|---|---|---|---|---|---|---|---|---|---|---|---|---|
| 4 | 280.992 | -35.4229 | RPJ 045403-693320 | T | 17.46 | 0.10 | 16.87 | 0.12 | --- | --- | 15.99 | 0.10 | 15.50 | 0.11 | --- | --- | --- | --- | 0.76 |
| 6 | 280.727 | -35.4454 | RPJ 045433-692035 | P | 14.36 | 0.04 | 13.59 | 0.02 | 12.41 | 0.03 | 9.95 | 0.03 | 8.94 | 0.05 | 8.02 | 0.03 | 6.65 | 0.03 | 0.66 |
| 14 | 277.724 | -35.8697 | RPJ 045733-665258 | T | --- | --- | --- | --- | --- | --- | 16.67 | 0.12 | 15.67 | 0.08 | 14.35 | 0.17 | 12.53 | 0.06 | 1.20 |
| 15 | 280.875 | -35.1052 | RPJ 045751-693333 | T | --- | --- | --- | --- | --- | --- | 16.74 | 0.11 | 15.96 | 0.11 | --- | --- | --- | --- | 0.88 |
| 16 | 280.499 | -35.1939 | RPJ 045754-691421 | T | --- | --- | --- | --- | --- | --- | --- | --- | 16.24 | 0.16 | --- | --- | --- | --- | 1.17 |
| 25 | 278.158 | -35.5662 | RPJ 045937-671805 | T | --- | --- | --- | --- | --- | --- | 16.37 | 0.14 | 16.48 | 0.14 | --- | --- | --- | --- | 1.69 |
| 27 | 281.325 | -34.7532 | RPJ 050032-700049 | P | 17.37 | 0.13 | 16.31 | 0.11 | 15.26 | 0.07 | 12.81 | 0.06 | 11.97 | 0.03 | 10.88 | 0.04 | 9.15 | 0.03 | 0.69 |
| 28 | 282.316 | -34.5041 | RPJ 050034-705200 | T | --- | --- | 15.22 | 0.05 | --- | --- | 10.45 | 0.03 | 9.60 | 0.02 | 8.97 | 0.03 | 8.31 | 0.02 | 0.74 |
| 30 | 279.879 | -35.0638 | RPJ 050052-684717 | T | --- | --- | --- | --- | --- | --- | 17.10 | 0.14 | --- | --- | --- | --- | --- | --- | 1.58 |
| 31 | 279.099 | -35.229 | RPJ 050058-680748 | T | --- | --- | --- | --- | --- | --- | 15.25 | 0.09 | 14.51 | 0.11 | 13.17 | 0.09 | 11.26 | 0.09 | 0.67 |
| 32 | 279.209 | -35.1891 | RPJ 050108-681337 | T | 16.27 | 0.06 | 15.70 | 0.07 | 15.66 | 0.11 | 15.71 | 0.11 | 15.85 | 0.14 | --- | --- | --- | --- | 1.91 |
| 37 | 280.284 | -34.8602 | RPJ 050203-690945 | T | --- | --- | --- | --- | --- | --- | 16.54 | 0.13 | 15.86 | 0.15 | --- | --- | 13.55 | 0.11 | 0.70 |
| 38 | 278.394 | -35.2731 | RPJ 050204-673343 | T | --- | --- | --- | --- | --- | --- | 16.55 | 0.09 | 15.73 | 0.11 | --- | --- | --- | --- | 1.23 |
| 39 | 280.166 | -34.8679 | RPJ 050216-690403 | T | --- | --- | --- | --- | --- | --- | 15.91 | 0.14 | --- | --- | --- | --- | --- | --- | 2.25 |
| 40 | 279.468 | -35.0006 | RPJ 050230-682848 | T | --- | --- | --- | --- | --- | --- | 16.42 | 0.10 | --- | --- | --- | --- | --- | --- | 0.28 |
| 43 | 281.383 | -34.5041 | RPJ 050311-700744 | T | 16.77 | 0.08 | 15.88 | 0.08 | 14.98 | 0.07 | 14.11 | 0.08 | 13.73 | 0.06 | 13.14 | 0.09 | 12.75 | 0.07 | 0.54 |
| 45 | 280.116 | -34.7408 | RPJ 050345-690341 | P | --- | --- | --- | --- | --- | --- | 14.28 | 0.10 | 13.37 | 0.05 | 12.58 | 0.07 | 11.55 | 0.06 | 0.44 |
| 47 | 279.99 | -34.7596 | RPJ 050351-685723 | T | 13.97 | 0.03 | 13.89 | 0.03 | 13.71 | 0.03 | 13.20 | 0.05 | 13.06 | 0.05 | 12.71 | 0.06 | --- | --- | 0.82 |
| 48 | 282.507 | -34.1438 | RPJ 050415-710717 | P | 17.58 | 0.10 | 16.87 | 0.13 | --- | --- | 16.75 | 0.08 | 16.71 | 0.18 | --- | --- | --- | --- | 0.68 |
| 50 | 278.417 | -35.0449 | RPJ 050421-673809 | T | 17.43 | 0.09 | 16.89 | 0.13 | 16.66 | 0.21 | 16.74 | 0.10 | 16.69 | 0.14 | --- | --- | --- | --- | 1.25 |
| 51 | 282.117 | -34.225 | RPJ 050424-704719 | P | --- | --- | --- | --- | --- | --- | 14.93 | 0.08 | 14.07 | 0.07 | 12.80 | 0.12 | --- | --- | 1.96 |
| 52 | 280.403 | -34.6064 | RPJ 050431-691928 | T | --- | --- | 16.40 | 0.10 | 14.25 | 0.04 | 11.76 | 0.05 | 10.97 | 0.03 | 10.26 | 0.04 | 9.53 | 0.04 | 1.07 |
| 53 | 278.691 | -34.9676 | RPJ 050434-675221 | T | 16.39 | 0.04 | 15.27 | 0.03 | 14.33 | 0.04 | 12.49 | 0.03 | 11.74 | 0.03 | 11.04 | 0.03 | 10.13 | 0.03 | 0.18 |
| 54 | 281.316 | -34.3962 | RPJ 050435-700618 | T | --- | --- | --- | --- | --- | --- | 15.05 | 0.17 | 14.67 | 0.08 | 12.39 | 0.14 | --- | --- | 0.31 |
| 56 | 277.45 | -35.2046 | RPJ 050440-664947 | P | 16.14 | 0.03 | 16.06 | 0.06 | 16.23 | 0.16 | 16.12 | 0.13 | --- | --- | --- | --- | --- | --- | 1.22 |

Table 3 (cont.)

| # | l deg | b deg | NAME | STAT | J mag | σ_J mag | H mag | σ_H mag | K_S mag | σ_KS mag | [3.6] mag | σ_3.6 mag | [4.5] mag | σ_4.5 mag | [5.8] mag | σ_5.8 mag | [8.0] mag | σ_8.0 mag | r arcs |
|---|---|---|---|---|---|---|---|---|---|---|---|---|---|---|---|---|---|---|---|
| 57 | 280.594 | -34.5498 | RPJ 050441-692925 | T | 16.31 | 0.05 | 15.60 | 0.05 | 15.47 | 0.08 | 15.28 | 0.05 | 15.27 | 0.08 | --- | --- | --- | --- | 2.44 |
| 60 | 282.023 | -34.2037 | RPJ 050454-704309 | T | 15.64 | 0.03 | 15.52 | 0.06 | 15.28 | 0.08 | 14.93 | 0.10 | 14.60 | 0.15 | 14.61 | 0.17 | --- | --- | 0.23 |
| 61 | 282.03 | -34.2019 | RPJ 050454-704332 | L | 13.92 | 0.03 | 13.81 | 0.03 | 13.61 | 0.03 | 13.15 | 0.06 | 12.89 | 0.04 | 12.67 | 0.06 | 12.32 | 0.08 | 0.06 |
| 62 | 279.674 | -34.7047 | RPJ 050510-684308 | T | --- | --- | --- | --- | --- | --- | --- | --- | 16.38 | 0.12 | --- | --- | --- | --- | 0.27 |
| 64 | 281.833 | -34.1661 | RPJ 050551-703443 | P | 14.89 | 0.03 | 14.31 | 0.03 | 14.04 | 0.03 | 13.63 | 0.04 | 13.56 | 0.06 | 13.41 | 0.11 | --- | --- | 0.70 |
| 67 | 281.959 | -34.093 | RPJ 050622-704155 | T | --- | --- | --- | --- | --- | --- | 16.61 | 0.11 | 16.45 | 0.13 | --- | --- | --- | --- | 1.65 |
| 68 | 279.666 | -34.588 | RPJ 050626-684430 | T | 16.53 | 0.06 | 15.98 | 0.09 | 15.74 | 0.11 | 15.61 | 0.07 | 15.67 | 0.09 | --- | --- | --- | --- | 0.71 |
| 69 | 279.664 | -34.5705 | RPJ 050638-684440 | L | --- | --- | 15.96 | 0.10 | --- | --- | 15.15 | 0.07 | --- | --- | --- | --- | --- | --- | 0.33 |
| 70 | 279.535 | -34.5873 | RPJ 050644-683813 | T | 14.73 | 0.03 | 14.03 | 0.03 | 13.78 | 0.03 | 13.69 | 0.05 | 13.75 | 0.05 | 13.51 | 0.11 | 13.27 | 0.10 | 0.49 |
| 73 | 280.217 | -34.4116 | RPJ 050706-691329 | T | --- | --- | --- | --- | --- | --- | --- | --- | 15.45 | 0.15 | --- | --- | --- | --- | 1.20 |
| 75 | 279.322 | -34.5509 | RPJ 050735-682832 | T | 16.58 | 0.06 | 16.48 | 0.13 | 16.18 | 0.17 | 16.04 | 0.11 | 15.68 | 0.11 | --- | --- | --- | --- | 0.17 |
| 77 | 279.129 | -34.5706 | RPJ 050747-681859 | T | --- | --- | --- | --- | --- | --- | 15.99 | 0.08 | 15.87 | 0.11 | --- | --- | --- | --- | 0.20 |
| 79 | 280.1 | -34.3432 | RPJ 050806-690854 | T | --- | --- | --- | --- | --- | --- | 15.96 | 0.05 | 14.83 | 0.05 | 13.50 | 0.06 | 11.62 | 0.04 | 1.36 |
| 83 | 280.343 | -34.2313 | RPJ 050847-692212 | T | --- | --- | --- | --- | --- | --- | 16.78 | 0.16 | --- | --- | --- | --- | --- | --- | 0.65 |
| 84 | 279.595 | -34.3804 | RPJ 050849-684404 | T | --- | --- | --- | --- | --- | --- | 15.67 | 0.11 | 15.11 | 0.13 | --- | --- | --- | --- | 1.29 |
| 85 | 279.909 | -34.3118 | RPJ 050852-690010 | T | --- | --- | --- | --- | --- | --- | 16.39 | 0.11 | --- | --- | --- | --- | --- | --- | 0.65 |
| 86 | 279.726 | -34.3329 | RPJ 050902-685103 | T | --- | --- | --- | --- | --- | --- | 16.72 | 0.08 | 15.56 | 0.06 | 14.41 | 0.11 | 12.66 | 0.07 | 0.51 |
| 88 | 278.206 | -34.6126 | RPJ 050911-673401 | P | --- | --- | --- | --- | --- | --- | 15.84 | 0.11 | 15.23 | 0.13 | --- | --- | --- | --- | 0.78 |
| 90 | 280.321 | -34.1931 | RPJ 050915-692143 | T | --- | --- | --- | --- | --- | --- | --- | --- | 15.74 | 0.10 | --- | --- | --- | --- | 0.26 |
| 93 | 279.97 | -34.214 | RPJ 050948-690427 | T | --- | --- | --- | --- | --- | --- | 16.41 | 0.11 | 16.43 | 0.17 | --- | --- | --- | --- | 2.08 |
| 94 | 279.432 | -34.3154 | RPJ 050951-683708 | T | --- | --- | --- | --- | --- | --- | 16.77 | 0.15 | --- | --- | --- | --- | --- | --- | 2.21 |
| 95 | 281.895 | -33.7884 | RPJ 051006-704335 | T | 16.95 | 0.07 | 16.50 | 0.11 | --- | --- | 16.51 | 0.07 | 16.46 | 0.10 | --- | --- | --- | --- | 2.79 |
| 98 | 280.949 | -33.9563 | RPJ 051028-695525 | T | 16.79 | 0.08 | 16.39 | 0.11 | --- | --- | 16.17 | 0.10 | 16.23 | 0.15 | --- | --- | --- | --- | 1.04 |
| 99 | 281.76 | -33.7409 | RPJ 051059-703745 | T | --- | --- | --- | --- | --- | --- | 16.88 | 0.11 | --- | --- | --- | --- | --- | --- | 1.66 |
| 102 | 279.939 | -34.0427 | RPJ 051144-690520 | T | --- | --- | --- | --- | --- | --- | 16.78 | 0.10 | 16.43 | 0.12 | --- | --- | --- | --- | 2.71 |
| 104 | 280.434 | -33.9442 | RPJ 051146-693040 | T | 16.35 | 0.06 | 15.77 | 0.07 | 15.49 | 0.10 | 15.35 | 0.08 | 15.19 | 0.12 | 14.36 | 0.15 | --- | --- | 0.72 |

Table 3 (cont.)

| # | l | b | NAME | STAT | J | $\sigma_J$ | H | $\sigma_H$ | $K_S$ | $\sigma_{K_S}$ | [3.6] | $\sigma_{3.6}$ | [4.5] | $\sigma_{4.5}$ | [5.8] | $\sigma_{5.8}$ | [8.0] | $\sigma_{8.0}$ | r |
|---|---|---|---|---|---|---|---|---|---|---|---|---|---|---|---|---|---|---|---|
| | deg | deg | | | mag | mag | mag | mag | mag | mag | mag | mag | mag | mag | mag | mag | mag | mag | arcs |
| 106 | 280.562 | -33.9021 | RPJ 051157-693729 | P | 14.72 | 0.02 | 14.38 | 0.03 | 14.18 | 0.04 | 13.66 | 0.06 | 13.38 | 0.05 | 13.25 | 0.09 | 11.91 | 0.08 | 2.36 |
| 109 | 279.91 | -33.9749 | RPJ 051232-690452 | T | --- | --- | --- | --- | --- | --- | 16.13 | 0.09 | --- | --- | --- | --- | --- | --- | 1.29 |
| 111 | 280.627 | -33.8101 | RPJ 051250-694152 | P | 16.67 | 0.06 | 16.32 | 0.11 | 15.93 | 0.14 | 14.67 | 0.06 | 14.22 | 0.08 | 13.43 | 0.11 | --- | --- | 0.69 |
| 114 | 280.387 | -33.7953 | RPJ 051331-693025 | T | 15.43 | 0.04 | 14.81 | 0.04 | 14.81 | 0.06 | 14.62 | 0.07 | 14.76 | 0.10 | --- | --- | --- | --- | 1.70 |
| 115 | 277.82 | -34.2186 | RPJ 051353-672017 | P | 15.13 | 0.06 | 13.83 | 0.05 | 12.59 | 0.04 | 10.51 | 0.06 | 9.81 | 0.04 | 8.60 | 0.04 | 7.32 | 0.08 | 1.81 |
| 118 | 281.688 | -33.4924 | RPJ 051404-703751 | T | 16.52 | 0.05 | 16.11 | 0.08 | 15.83 | 0.13 | 15.52 | 0.11 | 15.23 | 0.14 | --- | --- | --- | --- | 0.08 |
| 119 | 282.029 | -33.4133 | RPJ 051412-705533 | T | 17.18 | 0.08 | 16.62 | 0.12 | 16.37 | 0.19 | 16.44 | 0.10 | 16.15 | 0.12 | --- | --- | --- | --- | 2.13 |
| 120 | 279.28 | -33.9204 | RPJ 051423-683456 | T | --- | --- | --- | --- | --- | --- | 16.51 | 0.10 | --- | --- | --- | --- | --- | --- | 2.32 |
| 121 | 280.136 | -33.7257 | RPJ 051449-691908 | T | 14.35 | 0.03 | 13.73 | 0.03 | 13.52 | 0.03 | 13.36 | 0.04 | 13.45 | 0.05 | 13.33 | 0.12 | --- | --- | 0.40 |
| 123 | 279.303 | -33.8512 | RPJ 051505-683655 | T | --- | --- | --- | --- | --- | --- | 16.34 | 0.11 | 15.83 | 0.08 | 13.58 | 0.11 | 11.78 | 0.06 | 0.53 |
| 125 | 278.691 | -33.9004 | RPJ 051540-680628 | L | 14.28 | 0.03 | 13.67 | 0.03 | 13.44 | 0.03 | 13.09 | 0.03 | 13.03 | 0.03 | 12.97 | 0.06 | 13.11 | 0.10 | 0.95 |
| 127 | 279.443 | -33.7487 | RPJ 051555-684503 | T | 16.28 | 0.05 | 15.62 | 0.05 | 15.49 | 0.09 | 15.32 | 0.06 | 15.34 | 0.09 | --- | --- | --- | --- | 2.09 |
| 128 | 280.134 | -33.6029 | RPJ 051611-692036 | T | 15.03 | 0.04 | 14.53 | 0.04 | 14.34 | 0.04 | 14.45 | 0.12 | 14.48 | 0.08 | --- | --- | --- | --- | 0.49 |
| 130 | 279.635 | -33.6439 | RPJ 051642-685541 | T | 15.55 | 0.03 | 14.83 | 0.03 | 14.68 | 0.05 | --- | --- | 14.56 | 0.07 | 14.28 | 0.16 | --- | --- | 0.86 |
| 131 | 278.734 | -33.7794 | RPJ 051652-681001 | L | 15.57 | 0.03 | 14.97 | 0.04 | 14.76 | 0.05 | 14.48 | 0.06 | 14.30 | 0.05 | --- | --- | --- | --- | 0.60 |
| 132 | 280.082 | -33.5466 | RPJ 051654-691845 | T | --- | --- | --- | --- | --- | --- | 15.74 | 0.10 | --- | --- | --- | --- | --- | --- | 1.54 |
| 134 | 279.448 | -33.6218 | RPJ 051717-684650 | T | --- | --- | --- | --- | --- | --- | 16.67 | 0.13 | 16.56 | 0.13 | --- | --- | --- | --- | 0.21 |
| 135 | 279.444 | -33.6195 | RPJ 051719-684640 | P | 15.34 | 0.03 | 14.83 | 0.03 | 14.51 | 0.04 | 14.26 | 0.05 | 14.20 | 0.06 | 14.19 | 0.14 | --- | --- | 0.61 |
| 136 | 281.673 | -33.1812 | RPJ 051746-704119 | T | 17.59 | 0.12 | 17.03 | 0.15 | --- | --- | 16.73 | 0.09 | --- | --- | --- | --- | --- | --- | 0.71 |
| 138 | 280.006 | -33.4387 | RPJ 051815-691621 | P | --- | --- | --- | --- | --- | --- | 15.55 | 0.08 | 15.15 | 0.07 | 13.23 | 0.09 | 11.54 | 0.07 | 0.58 |
| 141 | 277.253 | -33.8409 | RPJ 051837-665645 | T | --- | --- | --- | --- | --- | --- | --- | --- | 14.18 | 0.06 | 12.85 | 0.06 | 11.16 | 0.04 | 0.44 |
| 143 | 280.375 | -33.3149 | RPJ 051855-693559 | T | --- | --- | --- | --- | --- | --- | 15.73 | 0.07 | 15.52 | 0.10 | --- | --- | --- | --- | 0.70 |
| 144 | 281.382 | -33.1298 | RPJ 051900-702744 | L | --- | --- | --- | --- | --- | --- | 17.00 | 0.09 | --- | --- | --- | --- | --- | --- | 2.85 |
| 148 | 281.472 | -33.0911 | RPJ 051916-703237 | T | 17.35 | 0.09 | 16.96 | 0.13 | 16.63 | 0.23 | 16.40 | 0.06 | 16.01 | 0.08 | --- | --- | --- | --- | 0.36 |
| 153 | 281.583 | -33.0139 | RPJ 051956-703903 | T | 16.68 | 0.06 | 16.05 | 0.07 | 15.84 | 0.11 | 15.52 | 0.05 | 15.46 | 0.09 | --- | --- | --- | --- | 0.51 |

Table 3 (cont.)

| # | l | b | NAME | STAT | J | $\sigma_J$ | H | $\sigma_H$ | $K_S$ | $\sigma_{KS}$ | [3.6] | $\sigma_{3.6}$ | [4.5] | $\sigma_{4.5}$ | [5.8] | $\sigma_{5.8}$ | [8.0] | $\sigma_{8.0}$ | r |
|---|---|---|---|---|---|---|---|---|---|---|---|---|---|---|---|---|---|---|---|
| | deg | deg | | | mag | mag | mag | mag | mag | mag | mag | mag | mag | mag | mag | mag | mag | mag | arcs |
| 154 | 280.557 | -33.1915 | RPJ 051958-694623 | T | --- | --- | --- | --- | 15.00 | 0.07 | 13.64 | 0.05 | 13.06 | 0.05 | 12.56 | 0.10 | 12.02 | 0.08 | 0.55 |
| 156 | 280.436 | -33.1915 | RPJ 052011-694028 | T | 10.58 | 0.03 | 9.64 | 0.03 | 9.08 | 0.03 | 8.17 | 0.05 | 8.02 | 0.04 | 7.76 | 0.03 | 7.53 | 0.03 | 1.76 |
| 157 | 279.353 | -33.3613 | RPJ 052016-684510 | L | 15.38 | 0.04 | 14.67 | 0.04 | 14.54 | 0.04 | 14.13 | 0.06 | 14.18 | 0.07 | 14.11 | 0.15 | --- | --- | 0.09 |
| 158 | 280.701 | -33.1382 | RPJ 052017-695407 | T | --- | --- | --- | --- | --- | --- | 16.53 | 0.12 | 16.48 | 0.11 | --- | --- | --- | --- | 1.44 |
| 159 | 279.139 | -33.39 | RPJ 052020-683421 | T | --- | --- | --- | --- | --- | --- | 16.13 | 0.09 | 15.29 | 0.09 | --- | --- | 12.51 | 0.07 | 0.58 |
| 160 | 279.153 | -33.3651 | RPJ 052034-683518 | P | 16.47 | 0.05 | 15.79 | 0.05 | 14.60 | 0.04 | 12.94 | 0.04 | 12.21 | 0.04 | 11.19 | 0.03 | 9.49 | 0.03 | 0.31 |
| 161 | 279.422 | -33.3213 | RPJ 052035-684901 | L | --- | --- | --- | --- | --- | --- | 17.01 | 0.11 | 16.68 | 0.15 | --- | --- | --- | --- | 0.51 |
| 163 | 281.052 | -33.0313 | RPJ 052049-701241 | L | 15.94 | 0.05 | 15.83 | 0.07 | 15.78 | 0.13 | 15.72 | 0.07 | 15.79 | 0.12 | --- | --- | --- | --- | 2.79 |
| 164 | 280.609 | -33.1025 | RPJ 052052-695002 | T | --- | --- | --- | --- | --- | --- | 15.55 | 0.10 | --- | --- | --- | --- | --- | --- | 1.25 |
| 166 | 280.589 | -33.094 | RPJ 052100-694907 | T | 14.99 | 0.03 | 14.28 | 0.03 | 14.16 | 0.03 | 14.07 | 0.06 | 14.07 | 0.07 | 13.83 | 0.13 | --- | --- | 1.68 |
| 167 | 280.335 | -33.1159 | RPJ 052114-693621 | T | --- | --- | --- | --- | --- | --- | --- | --- | 16.45 | 0.16 | --- | --- | --- | --- | 0.91 |
| 168 | 280.947 | -33.0138 | RPJ 052114-700745 | T | --- | --- | --- | --- | --- | --- | 16.85 | 0.13 | 16.52 | 0.13 | --- | --- | --- | --- | 1.57 |
| 173 | 278.267 | -33.4125 | RPJ 052129-675106 | T | 13.00 | 0.03 | 12.02 | 0.03 | 10.46 | 0.03 | 7.99 | 0.04 | 6.97 | 0.04 | 5.89 | 0.04 | 4.15 | 0.04 | 0.33 |
| 174 | 281.681 | -32.8609 | RPJ 052133-704548 | T | 13.34 | 0.03 | 12.44 | 0.03 | 12.18 | 0.03 | 11.97 | 0.05 | 12.14 | 0.04 | 11.94 | 0.07 | 11.85 | 0.08 | 2.56 |
| 175 | 277.247 | -33.5365 | RPJ 052141-665931 | T | --- | --- | --- | --- | --- | --- | 14.54 | 0.06 | 13.60 | 0.05 | 12.62 | 0.07 | 11.08 | 0.03 | 0.73 |
| 176 | 280.459 | -33.0467 | RPJ 052147-694315 | T | 15.82 | 0.05 | 15.69 | 0.08 | 15.51 | 0.09 | 15.14 | 0.12 | 14.96 | 0.09 | --- | --- | --- | --- | 0.60 |
| 177 | 280.464 | -33.0408 | RPJ 052150-694334 | P | --- | --- | --- | --- | --- | --- | 16.29 | 0.12 | --- | --- | --- | --- | --- | --- | 1.09 |
| 178 | 279.821 | -33.1408 | RPJ 052152-691043 | T | --- | --- | --- | --- | --- | --- | 14.52 | 0.05 | 13.77 | 0.07 | 13.05 | 0.06 | 12.27 | 0.05 | 0.27 |
| 179 | 279.59 | -33.1682 | RPJ 052158-685859 | T | 14.63 | 0.03 | 13.77 | 0.03 | 13.62 | 0.03 | 13.54 | 0.04 | 13.62 | 0.05 | 13.67 | 0.11 | --- | --- | 1.77 |
| 180 | 280.582 | -32.9908 | RPJ 052212-694957 | P | 13.44 | 0.02 | 13.13 | 0.03 | 13.10 | 0.03 | 13.02 | 0.07 | 13.06 | 0.05 | 12.99 | 0.16 | --- | --- | 1.14 |
| 183 | 278.478 | -33.3042 | RPJ 052218-680239 | T | 16.23 | 0.05 | 15.74 | 0.06 | 15.27 | 0.08 | 14.51 | 0.14 | --- | --- | --- | --- | --- | --- | 0.37 |
| 185 | 280.779 | -32.9451 | RPJ 052221-700013 | T | 14.51 | 0.03 | 13.73 | 0.03 | 13.55 | 0.03 | 13.45 | 0.06 | 13.46 | 0.06 | 13.48 | 0.08 | --- | --- | 2.42 |
| 186 | 281.434 | -32.8319 | RPJ 052223-703355 | T | 14.99 | 0.03 | 14.39 | 0.03 | 14.38 | 0.05 | 14.36 | 0.13 | 14.17 | 0.12 | 14.09 | 0.16 | --- | --- | 1.05 |
| 187 | 281.495 | -32.818 | RPJ 052226-703706 | P | --- | --- | --- | --- | --- | --- | 16.37 | 0.07 | 15.99 | 0.08 | --- | --- | 13.37 | 0.09 | 0.64 |
| 188 | 276.971 | -33.4902 | RPJ 052231-664621 | L | --- | --- | --- | --- | --- | --- | --- | --- | 14.49 | 0.08 | --- | --- | --- | --- | 1.44 |

Table 3 (cont.)

| # | l deg | b deg | NAME | STAT | J mag | σ_J mag | H mag | σ_H mag | K_S mag | σ_KS mag | [3.6] mag | σ_3.6 mag | [4.5] mag | σ_4.5 mag | [5.8] mag | σ_5.8 mag | [8.0] mag | σ_8.0 mag | r arcs |
|---|---|---|---|---|---|---|---|---|---|---|---|---|---|---|---|---|---|---|---|
| 190 | 278.324 | -33.2939 | RPJ 052238-675508 | T | 15.05 | 0.03 | 14.87 | 0.04 | 14.75 | 0.06 | 14.49 | 0.09 | 14.33 | 0.14 | --- | --- | --- | --- | 0.26 |
| 191 | 280.347 | -32.9817 | RPJ 052243-693828 | T | 12.60 | 0.03 | 11.33 | 0.03 | 10.20 | 0.02 | 8.81 | 0.05 | 8.16 | 0.03 | 7.55 | 0.03 | 6.96 | 0.02 | 0.29 |
| 192 | 276.859 | -33.4751 | RPJ 052249-664055 | P | 14.74 | 0.03 | 14.11 | 0.04 | 13.23 | 0.04 | --- | --- | --- | --- | 8.84 | 0.15 | --- | --- | 0.85 |
| 193 | 279.163 | -33.1504 | RPJ 052252-683805 | T | --- | --- | --- | --- | --- | --- | 17.28 | 0.13 | --- | --- | --- | --- | --- | --- | 0.25 |
| 194 | 280.385 | -32.9605 | RPJ 052254-694036 | L | 13.17 | 0.02 | 13.08 | 0.04 | 13.02 | 0.04 | 12.89 | 0.05 | 12.84 | 0.05 | 12.90 | 0.10 | 12.71 | 0.10 | 1.53 |
| 195 | 280.371 | -32.9438 | RPJ 052307-694004 | T | --- | --- | --- | --- | --- | --- | 15.76 | 0.09 | 15.77 | 0.16 | --- | --- | --- | --- | 2.17 |
| 196 | 280.553 | -32.9086 | RPJ 052311-694929 | T | --- | --- | --- | --- | --- | --- | 15.27 | 0.12 | 15.48 | 0.14 | --- | --- | --- | --- | 0.20 |
| 197 | 281.01 | -32.8201 | RPJ 052321-701305 | T | 16.37 | 0.08 | 15.81 | 0.08 | 15.60 | 0.11 | 15.56 | 0.08 | 15.63 | 0.11 | --- | --- | --- | --- | 1.28 |
| 198 | 280.722 | -32.8504 | RPJ 052332-695831 | P | 12.07 | 0.03 | 11.17 | 0.03 | 10.78 | 0.02 | 10.44 | 0.04 | 10.68 | 0.03 | 10.46 | 0.03 | 10.39 | 0.04 | 0.94 |
| 199 | 280.026 | -32.9294 | RPJ 052352-692309 | T | 16.47 | 0.08 | 15.93 | 0.09 | --- | --- | 15.51 | 0.09 | 15.56 | 0.10 | --- | --- | --- | --- | 1.77 |
| 203 | 279.955 | -32.9148 | RPJ 052409-691947 | T | 12.63 | 0.02 | 11.59 | 0.03 | 11.18 | 0.03 | 10.90 | 0.04 | 10.88 | 0.03 | 10.67 | 0.04 | 10.38 | 0.07 | 0.97 |
| 209 | 277.312 | -33.2654 | RPJ 052421-670516 | T | --- | --- | --- | --- | --- | --- | 16.31 | 0.08 | 16.15 | 0.10 | --- | --- | --- | --- | 0.58 |
| 211 | 279.563 | -32.9385 | RPJ 052432-690006 | L | 15.01 | 0.04 | 14.44 | 0.04 | 14.19 | 0.04 | 13.77 | 0.08 | 13.82 | 0.08 | 13.72 | 0.13 | --- | --- | 0.25 |
| 212 | 279.642 | -32.9179 | RPJ 052438-690413 | T | 15.30 | 0.04 | 15.01 | 0.05 | 14.86 | 0.06 | 14.82 | 0.11 | 14.76 | 0.06 | --- | --- | --- | --- | 0.19 |
| 213 | 280.933 | -32.7058 | RPJ 052449-701034 | T | --- | --- | --- | --- | --- | --- | 15.93 | 0.15 | 15.93 | 0.15 | --- | --- | --- | --- | 1.41 |
| 214 | 279.613 | -32.8923 | RPJ 052458-690304 | P | 15.17 | 0.03 | 15.08 | 0.05 | 14.85 | 0.06 | 14.32 | 0.06 | 14.15 | 0.05 | 13.88 | 0.14 | --- | --- | 0.26 |
| 216 | 280.5 | -32.7519 | RPJ 052504-694833 | T | --- | --- | --- | --- | --- | --- | 15.41 | 0.11 | 15.55 | 0.09 | --- | --- | --- | --- | 1.29 |
| 217 | 280.521 | -32.734 | RPJ 052514-694949 | T | 15.65 | 0.04 | 15.02 | 0.06 | 14.94 | 0.07 | 14.64 | 0.08 | 14.89 | 0.10 | --- | --- | --- | --- | 0.54 |
| 218 | 279.301 | -32.8717 | RPJ 052541-684746 | T | --- | --- | --- | --- | --- | --- | --- | --- | 16.35 | 0.15 | --- | --- | --- | --- | 1.98 |
| 219 | 277.657 | -33.0639 | RPJ 052558-672413 | T | 13.79 | 0.02 | 13.85 | 0.03 | 13.92 | 0.03 | 13.83 | 0.04 | 13.90 | 0.06 | 13.82 | 0.10 | --- | --- | 0.96 |
| 220 | 281.016 | -32.5762 | RPJ 052611-701605 | T | --- | --- | --- | --- | --- | --- | 15.59 | 0.06 | 15.16 | 0.08 | --- | --- | --- | --- | 0.06 |
| 221 | 281.848 | -32.4436 | RPJ 052612-705855 | T | --- | --- | --- | --- | --- | --- | 16.26 | 0.09 | 16.09 | 0.13 | --- | --- | --- | --- | 1.81 |
| 222 | 280.034 | -32.7186 | RPJ 052613-692545 | L | --- | --- | --- | --- | --- | --- | 17.27 | 0.12 | 17.33 | 0.17 | --- | --- | --- | --- | 0.41 |
| 224 | 280.852 | -32.5799 | RPJ 052625-700754 | T | --- | --- | --- | --- | 15.21 | 0.08 | 12.59 | 0.05 | 11.68 | 0.03 | 10.95 | 0.04 | 10.06 | 0.03 | 0.46 |
| 225 | 280.855 | -32.5789 | RPJ 052626-700805 | T | --- | --- | --- | --- | --- | --- | 16.29 | 0.12 | --- | --- | --- | --- | --- | --- | 1.77 |

Table 3 (cont.)

| # | l deg | b deg | NAME | STAT | J mag | σ_J mag | H mag | σ_H mag | K_S mag | σ_KS mag | [3.6] mag | σ_3.6 mag | [4.5] mag | σ_4.5 mag | [5.8] mag | σ_5.8 mag | [8.0] mag | σ_8.0 mag | r arcs |
|---|---|---|---|---|---|---|---|---|---|---|---|---|---|---|---|---|---|---|---|
| 228 | 281.261 | -32.501 | RPJ 052637-702907 | T | --- | --- | --- | --- | --- | --- | 13.53 | 0.05 | 12.58 | 0.03 | 11.73 | 0.04 | 10.70 | 0.03 | 1.28 |
| 229 | 281.87 | -32.4037 | RPJ 052639-710027 | T | 16.55 | 0.08 | 16.14 | 0.11 | --- | --- | 15.85 | 0.09 | 15.75 | 0.14 | --- | --- | --- | --- | 2.82 |
| 230 | 278.055 | -32.9426 | RPJ 052642-674505 | P | --- | --- | --- | --- | --- | --- | 15.49 | 0.08 | 15.00 | 0.10 | --- | --- | --- | --- | 0.46 |
| 232 | 278.068 | -32.928 | RPJ 052650-674554 | P | 17.16 | 0.08 | 16.49 | 0.12 | 15.73 | 0.11 | 13.65 | 0.04 | 12.87 | 0.04 | 12.02 | 0.06 | 11.11 | 0.07 | 0.90 |
| 233 | 279.77 | -32.689 | RPJ 052658-69125 | L | 16.24 | 0.06 | 15.69 | 0.07 | 15.51 | 0.10 | 15.33 | 0.08 | 15.35 | 0.11 | --- | --- | --- | --- | 0.11 |
| 234 | 281.198 | -32.4633 | RPJ 052711-702623 | T | --- | --- | --- | --- | --- | --- | 16.97 | 0.09 | --- | --- | --- | --- | --- | --- | 1.94 |
| 236 | 280.733 | -32.5181 | RPJ 052721-700235 | T | --- | --- | --- | --- | --- | --- | --- | --- | 16.16 | 0.11 | --- | --- | --- | --- | 1.02 |
| 237 | 278.743 | -32.782 | RPJ 052728-682052 | L | 15.79 | 0.04 | 15.15 | 0.05 | 15.06 | 0.07 | 14.56 | 0.05 | 14.48 | 0.07 | --- | --- | --- | --- | 1.21 |
| 242 | 279.495 | -32.627 | RPJ 052804-685947 | L | 15.75 | 0.05 | 14.77 | 0.06 | 13.47 | 0.03 | 11.06 | 0.08 | 10.12 | 0.05 | 9.29 | 0.11 | --- | --- | 0.59 |
| 244 | 280.484 | -32.4842 | RPJ 052809-695032 | T | --- | --- | --- | --- | --- | --- | 16.20 | 0.08 | 16.41 | 0.14 | --- | --- | --- | --- | 0.95 |
| 245 | 277.332 | -32.8847 | RPJ 052812-670935 | T | 16.23 | 0.05 | 15.64 | 0.06 | 15.53 | 0.10 | 15.54 | 0.08 | 15.56 | 0.11 | --- | --- | --- | --- | 2.80 |
| 247 | 280.433 | -32.4601 | RPJ 052830-694814 | T | --- | --- | --- | --- | --- | --- | 15.82 | 0.11 | 15.67 | 0.11 | --- | --- | --- | --- | 1.63 |
| 249 | 280.549 | -32.4244 | RPJ 052844-695422 | T | 14.24 | 0.03 | 13.43 | 0.04 | 13.30 | 0.04 | 13.07 | 0.04 | 13.19 | 0.04 | 13.19 | 0.09 | 13.02 | 0.09 | 0.39 |
| 250 | 277.03 | -32.8622 | RPJ 052846-665440 | T | 15.59 | 0.03 | 14.91 | 0.04 | 14.75 | 0.06 | 14.69 | 0.05 | 14.63 | 0.07 | 14.51 | 0.15 | --- | --- | 1.01 |
| 251 | 279.561 | -32.537 | RPJ 052858-690354 | P | --- | --- | 16.29 | 0.12 | --- | --- | 16.02 | 0.11 | 15.60 | 0.13 | --- | --- | --- | --- | 0.67 |
| 252 | 280.399 | -32.4148 | RPJ 052905-694657 | T | 16.07 | 0.06 | 15.59 | 0.07 | 15.44 | 0.10 | 15.26 | 0.06 | 15.18 | 0.06 | --- | --- | --- | --- | 0.88 |
| 253 | 280.348 | -32.409 | RPJ 052913-694428 | T | --- | --- | --- | --- | --- | --- | 15.50 | 0.08 | 14.66 | 0.07 | 13.43 | 0.11 | 11.87 | 0.06 | 1.50 |
| 255 | 279.524 | -32.5073 | RPJ 052921-690220 | P | 14.58 | 0.03 | 13.75 | 0.03 | 13.58 | 0.03 | 13.50 | 0.06 | 13.58 | 0.06 | 13.41 | 0.12 | --- | --- | 2.00 |
| 257 | 280.363 | -32.3777 | RPJ 052933-694531 | T | --- | --- | --- | --- | --- | --- | 15.97 | 0.08 | 16.04 | 0.11 | --- | --- | --- | --- | 0.41 |
| 260 | 276.756 | -32.7942 | RPJ 052946-664130 | T | 16.95 | 0.07 | 16.59 | 0.12 | 16.42 | 0.21 | 16.43 | 0.10 | 16.40 | 0.11 | --- | --- | --- | --- | 2.92 |
| 262 | 276.756 | -32.7923 | RPJ 052947-664130 | L | 16.52 | 0.05 | 15.97 | 0.07 | 15.66 | 0.12 | 15.71 | 0.07 | 15.67 | 0.09 | --- | --- | --- | --- | 2.31 |
| 265 | 280.143 | -32.3553 | RPJ 053009-693445 | P | 14.26 | 0.03 | 13.48 | 0.03 | 13.33 | 0.03 | 13.27 | 0.05 | 13.29 | 0.05 | 13.21 | 0.06 | 13.13 | 0.08 | 0.73 |
| 267 | 279.481 | -32.4352 | RPJ 053013-690050 | T | --- | --- | --- | --- | --- | --- | 16.92 | 0.10 | 16.20 | 0.13 | --- | --- | --- | --- | 0.18 |
| 269 | 280.923 | -32.2255 | RPJ 053026-701501 | P | --- | --- | --- | --- | --- | --- | 16.68 | 0.13 | --- | --- | --- | --- | --- | --- | 2.89 |
| 270 | 277.063 | -32.684 | RPJ 053033-665741 | L | --- | --- | --- | --- | --- | --- | 16.16 | 0.07 | 15.76 | 0.11 | --- | --- | 12.15 | 0.11 | 0.57 |

Table 3 (cont.)

| # | l | b | NAME | STAT | J | σ_J | H | σ_H | K_S | σ_KS | [3.6] | σ_{3.6} | [4.5] | σ_{4.5} | [5.8] | σ_{5.8} | [8.0] | σ_{8.0} | r |
|---|---|---|---|---|---|---|---|---|---|---|---|---|---|---|---|---|---|---|---|
| | deg | deg | | | mag | mag | mag | mag | mag | mag | mag | mag | mag | mag | mag | mag | mag | mag | arcs |
| 271 | 280.923 | -32.2015 | RPJ 053043-701515 | T | --- | --- | --- | --- | --- | --- | 15.64 | 0.08 | 15.43 | 0.10 | --- | --- | --- | --- | 1.33 |
| 272 | 279.044 | -32.4423 | RPJ 053044-683853 | T | --- | --- | --- | --- | --- | --- | 17.56 | 0.13 | --- | --- | --- | --- | --- | --- | 0.56 |
| 274 | 277.498 | -32.6013 | RPJ 053055-672005 | P | 15.25 | 0.07 | 14.22 | 0.06 | 12.98 | 0.04 | 10.81 | 0.05 | 9.98 | 0.04 | 8.79 | 0.05 | 7.40 | 0.05 | 1.31 |
| 275 | 280.409 | -32.2498 | RPJ 053057-694900 | T | --- | --- | --- | --- | --- | --- | 15.40 | 0.05 | 15.51 | 0.06 | --- | --- | --- | --- | 0.12 |
| 276 | 278.978 | -32.4263 | RPJ 053059-683542 | T | 16.04 | 0.05 | 15.02 | 0.05 | 13.76 | 0.04 | --- | --- | 11.65 | 0.09 | 10.30 | 0.05 | 8.57 | 0.05 | 0.96 |
| 277 | 279.66 | -32.3327 | RPJ 053106-691042 | P | --- | --- | --- | --- | --- | --- | 14.87 | 0.12 | 14.81 | 0.13 | --- | --- | --- | --- | 0.44 |
| 278 | 278.818 | -32.4238 | RPJ 053113-682742 | P | 17.30 | 0.11 | 16.69 | 0.15 | 15.90 | 0.15 | 14.42 | 0.05 | 13.83 | 0.06 | 13.43 | 0.09 | 12.87 | 0.10 | 0.58 |
| 279 | 281.673 | -32.0461 | RPJ 053119-705422 | L | 15.65 | 0.03 | 15.11 | 0.05 | 14.84 | 0.06 | 14.51 | 0.06 | 14.44 | 0.05 | 14.20 | 0.13 | --- | --- | 0.97 |
| 280 | 278.891 | -32.4037 | RPJ 053120-683134 | P | --- | --- | --- | --- | --- | --- | 16.25 | 0.07 | 16.30 | 0.13 | --- | --- | --- | --- | 0.42 |
| 281 | 280.433 | -32.1999 | RPJ 053129-695042 | T | --- | --- | --- | --- | --- | --- | 17.14 | 0.13 | --- | --- | --- | --- | --- | --- | 0.38 |
| 283 | 282.043 | -31.9632 | RPJ 053141-711346 | L | 14.81 | 0.03 | 14.24 | 0.03 | 13.97 | 0.03 | 13.66 | 0.05 | 13.60 | 0.04 | 13.59 | 0.11 | --- | --- | 0.53 |
| 286 | 281.972 | -31.969 | RPJ 053145-711008 | P | --- | --- | --- | --- | --- | --- | 16.80 | 0.09 | 16.44 | 0.13 | --- | --- | --- | --- | 0.07 |
| 287 | 280.331 | -32.1874 | RPJ 053147-694541 | T | --- | --- | --- | --- | --- | --- | 16.65 | 0.10 | 16.52 | 0.14 | --- | --- | --- | --- | 0.45 |
| 290 | 281.967 | -31.9624 | RPJ 053150-710955 | P | 16.14 | 0.05 | 15.43 | 0.06 | 15.25 | 0.08 | 14.92 | 0.06 | 14.80 | 0.06 | --- | --- | --- | --- | 1.35 |
| 292 | 281.515 | -32.0149 | RPJ 053157-704646 | T | --- | --- | --- | --- | --- | --- | 17.08 | 0.15 | --- | --- | --- | --- | --- | --- | 0.24 |
| 293 | 279.032 | -32.3081 | RPJ 053212-683924 | P | 14.82 | 0.02 | 14.21 | 0.03 | 13.96 | 0.03 | 13.63 | 0.04 | 13.57 | 0.04 | 13.53 | 0.10 | --- | --- | 0.12 |
| 294 | 281.521 | -31.9718 | RPJ 053228-704728 | P | 15.36 | 0.03 | 14.73 | 0.04 | 14.48 | 0.05 | 14.25 | 0.06 | 14.20 | 0.06 | 14.09 | 0.14 | --- | --- | 0.79 |
| 297 | 279.915 | -32.1712 | RPJ 053233-692456 | L | --- | --- | --- | --- | --- | --- | 16.16 | 0.09 | 15.89 | 0.11 | --- | --- | --- | --- | 0.27 |
| 299 | 280.049 | -32.1466 | RPJ 053239-693152 | T | --- | --- | --- | --- | 14.78 | 0.05 | 12.41 | 0.06 | 11.60 | 0.03 | 10.63 | 0.05 | 9.15 | 0.05 | 0.99 |
| 300 | 280.028 | -32.1484 | RPJ 053239-693049 | T | --- | --- | 16.04 | 0.10 | 14.93 | 0.07 | 12.13 | 0.08 | 11.31 | 0.05 | 10.29 | 0.07 | 8.87 | 0.06 | 0.03 |
| 302 | 281.347 | -31.9752 | RPJ 053242-703840 | T | --- | --- | --- | --- | --- | --- | 16.15 | 0.09 | 16.29 | 0.14 | --- | --- | --- | --- | 1.75 |
| 304 | 281.318 | -31.9671 | RPJ 053251-703717 | T | --- | --- | --- | --- | --- | --- | 16.55 | 0.10 | --- | --- | --- | --- | --- | --- | 0.25 |
| 305 | 277.883 | -32.3727 | RPJ 053252-674108 | P | 15.42 | 0.04 | 14.47 | 0.04 | 13.25 | 0.03 | 10.87 | 0.07 | 10.07 | 0.04 | 8.75 | 0.05 | 7.19 | 0.06 | 0.36 |
| 306 | 280.974 | -32.0108 | RPJ 053252-701935 | T | --- | --- | --- | --- | --- | --- | 16.62 | 0.12 | --- | --- | --- | --- | --- | --- | 1.69 |
| 310 | 282.104 | -31.8368 | RPJ 053308-711803 | T | --- | --- | --- | --- | --- | --- | 14.90 | 0.06 | 14.23 | 0.07 | 13.52 | 0.11 | 11.79 | 0.07 | 0.06 |

Table 3 (cont.)

| # | $l$ | $b$ | NAME | STAT | J | $\sigma_J$ | H | $\sigma_H$ | $K_S$ | $\sigma_{KS}$ | [3.6] | $\sigma_{3.6}$ | [4.5] | $\sigma_{4.5}$ | [5.8] | $\sigma_{5.8}$ | [8.0] | $\sigma_{8.0}$ | r |
|---|---|---|---|---|---|---|---|---|---|---|---|---|---|---|---|---|---|---|---|
|   | deg | deg |   |   | mag | mag | mag | mag | mag | mag | mag | mag | mag | mag | mag | mag | mag | mag | arcs |
| 311 | 279.018 | -32.2016 | RPJ 053323-683933 | L | --- | --- | --- | --- | --- | --- | 16.49 | 0.11 | 16.17 | 0.12 | --- | --- | --- | --- | 0.19 |
| 315 | 277.555 | -32.3406 | RPJ 053333-672454 | L | 14.48 | 0.03 | 14.61 | 0.03 | 14.58 | 0.05 | 14.63 | 0.04 | 14.67 | 0.05 | 14.59 | 0.15 | --- | --- | 0.13 |
| 316 | 279.663 | -32.1022 | RPJ 053340-691250 | P | 15.45 | 0.03 | 15.07 | 0.05 | 14.79 | 0.05 | 14.29 | 0.05 | 14.09 | 0.05 | 14.03 | 0.14 | --- | --- | 0.69 |
| 317 | 279.915 | -31.9822 | RPJ 053441-692630 | T | 12.47 | 0.02 | 10.99 | 0.02 | 9.91 | 0.03 | 8.95 | 0.05 | 8.25 | 0.03 | 7.75 | 0.03 | 7.25 | 0.02 | 0.51 |
| 318 | 279.175 | -32.0551 | RPJ 053448-684835 | T | --- | --- | --- | --- | --- | --- | 15.14 | 0.07 | 14.45 | 0.06 | 13.14 | 0.09 | 11.24 | 0.05 | 0.50 |
| 320 | 280.544 | -31.8445 | RPJ 053525-695921 | T | 13.94 | 0.03 | 13.07 | 0.03 | 12.82 | 0.03 | 12.73 | 0.03 | 12.83 | 0.04 | 12.78 | 0.07 | 12.55 | 0.06 | 0.78 |
| 321 | 277.789 | -32.1302 | RPJ 053530-673805 | P | 17.00 | 0.08 | 16.64 | 0.14 | --- | --- | 16.55 | 0.15 | --- | --- | --- | --- | --- | --- | 0.48 |
| 322 | 276.871 | -32.1907 | RPJ 053544-665122 | P | 16.58 | 0.06 | 16.00 | 0.08 | 15.79 | 0.12 | 15.43 | 0.06 | 15.30 | 0.07 | --- | --- | --- | --- | 0.91 |
| 324 | 277.021 | -32.1682 | RPJ 053549-665905 | T | --- | --- | --- | --- | --- | --- | 17.39 | 0.12 | --- | --- | --- | --- | --- | --- | 0.44 |
| 325 | 279.397 | -31.9272 | RPJ 053556-690045 | P | --- | --- | --- | --- | 12.63 | 0.05 | 10.16 | 0.03 | 9.20 | 0.03 | 8.42 | 0.03 | 7.43 | 0.03 | 0.27 |
| 326 | 277.922 | -32.0663 | RPJ 053602-674516 | P | 16.88 | 0.07 | 15.81 | 0.07 | 14.60 | 0.05 | 12.87 | 0.04 | 12.17 | 0.04 | 11.41 | 0.05 | 10.29 | 0.05 | 0.79 |
| 327 | 282.045 | -31.5976 | RPJ 053611-711719 | T | --- | --- | --- | --- | --- | --- | 16.85 | 0.17 | --- | --- | --- | --- | --- | --- | 2.57 |
| 328 | 282.047 | -31.5958 | RPJ 053612-711724 | P | 14.11 | 0.02 | 13.79 | 0.03 | 13.80 | 0.03 | 13.76 | 0.06 | 13.68 | 0.07 | 13.67 | 0.16 | --- | --- | 2.98 |
| 329 | 279.309 | -31.9094 | RPJ 053614-685627 | P | --- | --- | --- | --- | 16.01 | 0.15 | 15.27 | 0.11 | 15.02 | 0.08 | --- | --- | --- | --- | 2.87 |
| 331 | 281.113 | -31.6823 | RPJ 053632-702925 | T | 16.69 | 0.08 | 16.00 | 0.10 | --- | --- | 15.82 | 0.07 | 15.96 | 0.10 | --- | --- | --- | --- | 2.75 |
| 332 | 279.811 | -31.8257 | RPJ 053635-692228 | L | --- | --- | --- | --- | --- | --- | 16.79 | 0.14 | 16.43 | 0.14 | --- | --- | --- | --- | 0.66 |
| 334 | 281.027 | -31.6847 | RPJ 053638-702505 | T | --- | --- | --- | --- | --- | --- | --- | --- | 16.91 | 0.14 | --- | --- | --- | --- | 0.82 |
| 335 | 279.803 | -31.8177 | RPJ 053641-692208 | P | 16.53 | 0.06 | 15.97 | 0.08 | 16.05 | 0.15 | 15.65 | 0.07 | 15.66 | 0.13 | --- | --- | --- | --- | 0.76 |
| 336 | 277.674 | -32.0221 | RPJ 053644-673259 | P | 15.55 | 0.04 | 15.00 | 0.04 | 14.92 | 0.06 | 14.51 | 0.05 | 14.46 | 0.05 | --- | --- | --- | --- | 0.41 |
| 337 | 279.891 | -31.7972 | RPJ 053648-692644 | P | --- | --- | --- | --- | --- | --- | 16.04 | 0.07 | 15.77 | 0.09 | --- | --- | --- | --- | 0.27 |
| 338 | 279.836 | -31.802 | RPJ 053649-692355 | L | 16.56 | 0.06 | 16.21 | 0.10 | --- | --- | 15.45 | 0.09 | 15.28 | 0.13 | --- | --- | --- | --- | 0.86 |
| 339 | 277.69 | -32.0014 | RPJ 053656-673356 | P | 16.82 | 0.08 | 16.13 | 0.11 | 15.69 | 0.13 | --- | --- | --- | --- | 12.08 | 0.17 | --- | --- | 0.71 |
| 340 | 279.786 | -31.7905 | RPJ 053700-692128 | T | 16.69 | 0.07 | 16.26 | 0.10 | 16.00 | 0.15 | 15.92 | 0.09 | 15.76 | 0.07 | --- | --- | --- | --- | 1.31 |
| 342 | 278.484 | -31.9155 | RPJ 053704-681442 | P | 17.01 | 0.09 | 16.06 | 0.10 | 15.27 | 0.09 | --- | --- | 12.88 | 0.14 | 10.86 | 0.16 | --- | --- | 1.17 |

Table 3 (cont.)

| # | l | b | NAME | STAT | J | $\sigma_J$ | H | $\sigma_H$ | $K_S$ | $\sigma_{KS}$ | [3.6] | $\sigma_{3.6}$ | [4.5] | $\sigma_{4.5}$ | [5.8] | $\sigma_{5.8}$ | [8.0] | $\sigma_{8.0}$ | r |
|---|---|---|---|---|---|---|---|---|---|---|---|---|---|---|---|---|---|---|---|
| | deg | deg | | | mag | mag | mag | mag | mag | mag | mag | mag | mag | mag | mag | mag | mag | mag | arcs |
| 344 | 279.895 | -31.7702 | RPJ 053706-692709 | P | --- | --- | --- | --- | --- | --- | 16.02 | 0.10 | 15.84 | 0.09 | --- | --- | --- | --- | 0.60 |
| 345 | 280.919 | -31.656 | RPJ 053707-701951 | T | --- | --- | --- | --- | --- | --- | 17.11 | 0.13 | --- | --- | --- | --- | --- | --- | 0.82 |
| 346 | 282.146 | -31.5057 | RPJ 053710-712313 | T | 17.58 | 0.15 | 16.58 | 0.15 | 15.43 | 0.10 | 13.77 | 0.04 | 13.00 | 0.04 | 12.35 | 0.06 | 11.35 | 0.04 | 0.74 |
| 347 | 277.285 | -32.0139 | RPJ 053710-671323 | L | 16.01 | 0.04 | 15.20 | 0.05 | 15.09 | 0.07 | 14.90 | 0.07 | 14.87 | 0.05 | --- | --- | --- | --- | 0.25 |
| 355 | 280.681 | -31.6518 | RPJ 053729-700750 | T | --- | --- | --- | --- | --- | --- | 16.31 | 0.10 | 16.74 | 0.14 | --- | --- | --- | --- | 0.44 |
| 356 | 280.85 | -31.6327 | RPJ 053729-701633 | T | --- | --- | --- | --- | --- | --- | 12.94 | 0.06 | 12.96 | 0.03 | 12.72 | 0.09 | 12.65 | 0.08 | 0.50 |
| 357 | 281.901 | -31.5066 | RPJ 053731-711048 | P | 13.36 | 0.03 | 12.48 | 0.03 | 12.17 | 0.03 | 12.03 | 0.05 | 12.13 | 0.03 | 12.03 | 0.06 | 11.97 | 0.05 | 1.49 |
| 358 | 277.249 | -31.9601 | RPJ 053745-671153 | T | 15.17 | 0.03 | 14.61 | 0.03 | 14.29 | 0.04 | 13.83 | 0.05 | 13.82 | 0.05 | 13.80 | 0.13 | --- | --- | 0.44 |
| 359 | 279.979 | -31.7025 | RPJ 053746-693153 | P | 13.33 | 0.03 | 12.38 | 0.03 | 12.16 | 0.03 | 12.00 | 0.06 | 12.16 | 0.04 | 12.04 | 0.06 | 11.94 | 0.06 | 1.76 |
| 360 | 278.967 | -31.8019 | RPJ 053748-683954 | P | 16.59 | 0.07 | 16.51 | 0.14 | 16.22 | 0.19 | 15.45 | 0.07 | 15.20 | 0.10 | --- | --- | --- | --- | 0.22 |
| 361 | 281.156 | -31.5403 | RPJ 053810-703245 | L | --- | --- | --- | --- | --- | --- | 16.86 | 0.13 | 16.59 | 0.14 | --- | --- | --- | --- | 2.24 |
| 364 | 279.99 | -31.6425 | RPJ 053826-693251 | P | 14.51 | 0.03 | 14.15 | 0.03 | 14.07 | 0.04 | 13.93 | 0.05 | 13.91 | 0.06 | --- | --- | --- | --- | 2.62 |
| 367 | 280.883 | -31.5279 | RPJ 053840-701901 | T | 17.02 | 0.09 | 16.20 | 0.10 | --- | --- | 15.96 | 0.08 | 15.93 | 0.11 | --- | --- | --- | --- | 1.58 |
| 368 | 280.01 | -31.6091 | RPJ 053848-693407 | P | --- | --- | --- | --- | --- | --- | 15.85 | 0.07 | 15.49 | 0.13 | --- | --- | --- | --- | 0.49 |
| 372 | 280.027 | -31.5861 | RPJ 053902-693509 | P | 16.70 | 0.07 | 16.08 | 0.08 | 15.96 | 0.14 | 15.92 | 0.12 | 15.61 | 0.12 | --- | --- | --- | --- | 2.74 |
| 373 | 280.33 | -31.5518 | RPJ 053905-695045 | L | --- | --- | --- | --- | --- | --- | 16.11 | 0.10 | 16.08 | 0.16 | --- | --- | --- | --- | 2.46 |
| 375 | 280.028 | -31.5794 | RPJ 053907-693514 | P | 15.59 | 0.04 | 15.55 | 0.06 | 15.24 | 0.08 | 15.33 | 0.12 | --- | --- | --- | --- | --- | --- | 0.24 |
| 376 | 278.863 | -31.6769 | RPJ 053916-683527 | T | 17.21 | 0.10 | 16.34 | 0.12 | 15.27 | 0.08 | 13.99 | 0.05 | 13.43 | 0.04 | 12.92 | 0.07 | 12.12 | 0.05 | 0.56 |
| 377 | 281.114 | -31.4442 | RPJ 053922-703124 | T | --- | --- | --- | --- | --- | --- | 16.77 | 0.11 | 16.18 | 0.11 | --- | --- | --- | --- | 1.45 |
| 378 | 279.318 | -31.6143 | RPJ 053930-685857 | P | --- | --- | --- | --- | --- | --- | 15.77 | 0.10 | 15.44 | 0.11 | --- | --- | --- | --- | 0.22 |
| 382 | 281.872 | -31.3325 | RPJ 053942-711044 | P | --- | --- | --- | --- | 14.95 | 0.07 | 11.95 | 0.06 | 11.07 | 0.05 | 10.09 | 0.04 | 8.88 | 0.03 | 0.95 |
| 384 | 281.837 | -31.3234 | RPJ 053952-710902 | T | 17.14 | 0.10 | 16.15 | 0.10 | 15.30 | 0.09 | 12.82 | 0.08 | 12.11 | 0.04 | 10.06 | 0.03 | 8.48 | 0.02 | 0.66 |
| 387 | 279.3 | -31.5581 | RPJ 054008-685826 | P | 16.25 | 0.05 | 16.07 | 0.09 | 15.94 | 0.16 | 15.12 | 0.07 | 14.64 | 0.08 | 13.96 | 0.13 | --- | --- | 0.52 |
| 388 | 278.732 | -31.5918 | RPJ 054019-682919 | P | --- | --- | --- | --- | --- | --- | 16.61 | 0.06 | 16.31 | 0.07 | --- | --- | --- | --- | 0.46 |
| 389 | 279.582 | -31.5142 | RPJ 054020-691300 | P | --- | --- | --- | --- | --- | --- | 15.40 | 0.07 | 15.31 | 0.07 | --- | --- | --- | --- | 0.76 |

Table 3 (cont.)

| # | l | b | NAME | STAT | J | $\sigma_J$ | H | $\sigma_H$ | $K_S$ | $\sigma_{KS}$ | [3.6] | $\sigma_{3.6}$ | [4.5] | $\sigma_{4.5}$ | [5.8] | $\sigma_{5.8}$ | [8.0] | $\sigma_{8.0}$ | r |
|---|---|---|---|---|---|---|---|---|---|---|---|---|---|---|---|---|---|---|---|
| | deg | deg | | | mag | mag | mag | mag | mag | mag | mag | mag | mag | mag | mag | mag | mag | mag | arcs |
| 390 | 281.58 | -31.3027 | RPJ 054028-705609 | P | 15.73 | 0.04 | 15.12 | 0.05 | 14.99 | 0.06 | 14.89 | 0.06 | 15.04 | 0.08 | --- | --- | --- | --- | 2.78 |
| 391 | 280.389 | -31.4255 | RPJ 054028-695439 | T | --- | --- | --- | --- | --- | --- | 16.33 | 0.10 | 16.02 | 0.12 | --- | --- | --- | --- | 0.43 |
| 392 | 281.125 | -31.3436 | RPJ 054033-703240 | L | --- | --- | --- | --- | --- | --- | 12.97 | 0.04 | 12.13 | 0.02 | 10.72 | 0.03 | 8.90 | 0.02 | 0.51 |
| 394 | 281.034 | -31.3366 | RPJ 054045-702806 | T | 15.11 | 0.03 | 14.49 | 0.04 | 13.53 | 0.04 | 11.00 | 0.05 | 10.05 | 0.05 | 8.88 | 0.05 | --- | --- | 0.96 |
| 395 | 281.362 | -31.291 | RPJ 054053-704508 | T | --- | --- | 16.71 | 0.14 | --- | --- | 16.43 | 0.11 | 16.30 | 0.14 | --- | --- | --- | --- | 2.86 |
| 396 | 278.931 | -31.5204 | RPJ 054055-683954 | P | --- | --- | --- | --- | --- | --- | 16.97 | 0.14 | 16.25 | 0.09 | --- | --- | --- | --- | 0.50 |
| 397 | 279.597 | -31.4611 | RPJ 054055-691409 | P | --- | --- | --- | --- | --- | --- | 15.55 | 0.08 | 15.33 | 0.09 | --- | --- | --- | --- | 0.58 |
| 401 | 281.654 | -31.2181 | RPJ 054124-710030 | P | --- | --- | --- | --- | --- | --- | 16.73 | 0.09 | --- | --- | --- | --- | --- | --- | 1.71 |
| 403 | 279.084 | -31.4595 | RPJ 054126-684802 | P | 14.97 | 0.03 | 14.38 | 0.03 | 14.11 | 0.03 | 13.73 | 0.07 | 13.64 | 0.04 | 13.67 | 0.11 | --- | --- | 1.38 |
| 405 | 281.025 | -31.2426 | RPJ 054152-702818 | T | 16.40 | 0.07 | 15.72 | 0.08 | 15.47 | 0.09 | 15.48 | 0.07 | 15.24 | 0.16 | --- | --- | --- | --- | 2.65 |
| 406 | 281.72 | -31.1706 | RPJ 054153-710415 | T | 15.09 | 0.03 | 14.46 | 0.03 | 14.33 | 0.04 | 14.21 | 0.05 | 14.29 | 0.04 | 14.21 | 0.15 | --- | --- | 2.94 |
| 407 | 280.959 | -31.2361 | RPJ 054202-702459 | T | --- | --- | --- | --- | --- | --- | 17.58 | 0.16 | --- | --- | --- | --- | --- | --- | 0.79 |
| 414 | 281.03 | -31.1811 | RPJ 054236-702859 | T | --- | --- | --- | --- | --- | --- | 17.13 | 0.14 | --- | --- | --- | --- | --- | --- | 2.06 |
| 416 | 281.113 | -31.1372 | RPJ 054302-703332 | T | 16.63 | 0.07 | 16.08 | 0.10 | 15.79 | 0.13 | 15.86 | 0.06 | 15.85 | 0.10 | --- | --- | --- | --- | 2.93 |
| 417 | 277.955 | -31.3836 | RPJ 054312-675053 | P | 15.92 | 0.04 | 15.96 | 0.09 | --- | --- | 16.08 | 0.07 | 15.96 | 0.09 | --- | --- | --- | --- | 0.28 |
| 418 | 280.4 | -31.1811 | RPJ 054317-695651 | T | --- | --- | --- | --- | --- | --- | 15.72 | 0.16 | --- | --- | --- | --- | --- | --- | 0.42 |
| 420 | 279.776 | -31.2165 | RPJ 054330-692446 | P | 16.14 | 0.11 | 14.78 | 0.10 | 12.71 | 0.03 | 9.94 | 0.03 | 8.79 | 0.05 | 7.81 | 0.03 | 6.46 | 0.03 | 0.17 |
| 422 | 281.116 | -31.0858 | RPJ 054338-703401 | P | --- | --- | --- | --- | --- | --- | 16.28 | 0.11 | --- | --- | --- | --- | --- | --- | 0.16 |
| 425 | 279.625 | -31.1616 | RPJ 054415-691722 | P | 16.72 | 0.06 | 16.48 | 0.11 | --- | --- | 16.01 | 0.14 | --- | --- | --- | --- | --- | --- | 0.39 |
| 429 | 279.611 | -31.1486 | RPJ 054425-691642 | P | 15.43 | 0.03 | 15.16 | 0.04 | 14.96 | 0.06 | 14.33 | 0.06 | 14.05 | 0.06 | --- | --- | --- | --- | 0.15 |
| 433 | 280.039 | -31.081 | RPJ 054448-693859 | P | 16.22 | 0.05 | 15.93 | 0.08 | 15.86 | 0.13 | 15.75 | 0.11 | 15.64 | 0.08 | --- | --- | --- | --- | 0.24 |
| 434 | 279.637 | -31.0951 | RPJ 054500-691819 | P | 17.39 | 0.12 | 15.72 | 0.07 | 13.70 | 0.03 | 10.72 | 0.03 | 9.67 | 0.03 | 8.82 | 0.04 | 7.86 | 0.03 | 0.33 |
| 435 | 277.544 | -31.237 | RPJ 054501-673034 | T | --- | --- | --- | --- | --- | --- | 17.39 | 0.14 | --- | --- | --- | --- | --- | --- | 2.73 |

Table 3 (cont.)

| # | *l* deg | *b* deg | NAME | STAT | J mag | σ_J mag | H mag | σ_H mag | K_S mag | σ_KS mag | [3.6] mag | σ_3.6 mag | [4.5] mag | σ_4.5 mag | [5.8] mag | σ_5.8 mag | [8.0] mag | σ_8.0 mag | r arcs |
|---|---|---|---|---|---|---|---|---|---|---|---|---|---|---|---|---|---|---|---|
| 436 | 279.478 | -31.0911 | RPJ 054511-691013 | P | --- | --- | --- | --- | --- | --- | 16.07 | 0.05 | 15.76 | 0.07 | --- | --- | --- | --- | 0.23 |
| 437 | 281.585 | -30.9131 | RPJ 054512-705905 | T | --- | --- | --- | --- | --- | --- | 16.93 | 0.13 | --- | --- | --- | --- | --- | --- | 2.06 |
| 438 | 280.028 | -31.0428 | RPJ 054515-693837 | P | --- | --- | --- | --- | --- | --- | 16.81 | 0.13 | --- | --- | --- | --- | --- | --- | 1.21 |
| 439 | 280.186 | -31.0289 | RPJ 054515-694648 | L | 13.13 | 0.03 | 11.76 | 0.03 | 10.73 | 0.02 | 9.03 | 0.04 | 8.50 | 0.03 | 8.09 | 0.02 | 7.78 | 0.02 | 0.35 |
| 441 | 281.272 | -30.9115 | RPJ 054534-704303 | T | --- | --- | --- | --- | --- | --- | 16.39 | 0.10 | 16.23 | 0.13 | --- | --- | --- | --- | 2.47 |
| 443 | 279.447 | -31.0233 | RPJ 054558-690857 | P | 14.76 | 0.03 | 14.42 | 0.03 | 14.22 | 0.04 | 13.83 | 0.08 | 13.69 | 0.05 | 13.41 | 0.10 | --- | --- | 0.60 |
| 446 | 279.958 | -30.9518 | RPJ 054621-693531 | T | --- | --- | --- | --- | --- | --- | 16.48 | 0.09 | --- | --- | --- | --- | --- | --- | 2.41 |
| 447 | 280.932 | -30.8711 | RPJ 054623-702555 | T | --- | --- | --- | --- | --- | --- | 16.40 | 0.10 | 16.29 | 0.14 | --- | --- | --- | --- | 0.11 |
| 449 | 281.527 | -30.7996 | RPJ 054639-705650 | T | --- | --- | --- | --- | --- | --- | 16.13 | 0.14 | 15.94 | 0.15 | 14.06 | 0.12 | 12.64 | 0.06 | 0.67 |
| 451 | 280.98 | -30.8145 | RPJ 054701-702842 | T | --- | --- | --- | --- | --- | --- | 16.98 | 0.16 | 16.39 | 0.15 | --- | --- | --- | --- | 2.60 |
| 452 | 280.813 | -30.7897 | RPJ 054728-702015 | T | --- | --- | --- | --- | --- | --- | 16.69 | 0.09 | 16.25 | 0.11 | --- | --- | --- | --- | 1.61 |
| 454 | 278.07 | -30.8898 | RPJ 054822-675853 | P | 14.57 | 0.03 | 13.09 | 0.04 | 11.73 | 0.02 | 10.20 | 0.04 | 9.45 | 0.03 | 8.80 | 0.03 | 8.02 | 0.03 | 0.25 |
| 456 | 280.95 | -30.6225 | RPJ 054920-702809 | T | 16.16 | 0.04 | 15.69 | 0.06 | 15.62 | 0.12 | 15.35 | 0.07 | 15.42 | 0.08 | --- | --- | --- | --- | 1.47 |
| 458 | 280.574 | -30.6029 | RPJ 054953-700855 | T | --- | --- | --- | --- | --- | --- | 16.95 | 0.14 | 16.69 | 0.14 | --- | --- | --- | --- | 0.64 |

# Figure Captions

**Figure 1**

The distributions of LMC nebulae (small blue disks) and Galactic bulge sources (large red disks) in the [3.6]/([3.6]-[8.0] magnitude-colour plane, where the magnitudes of LMC sources have been reduced to the value they would have in the Galactic centre. The dashed lines indicate the SAGE completeness limits for the LMC nebulae. As would be expected, there are relatively few LMC nebulae below the two dashed completeness limits, in a region where most of the GBPNe appear to be located. This suggests that most of the LMC PNe have yet to be detected.

**Figure 2**

As for Fig. 1, but for the [8.0]/([3.6]-[8.0]) plane.

**Figure 3**

The variation in normalised source numbers with 3.6 and 8.0 $\mu$m magnitudes. Notice how the 3.6 $\mu$m bulge and LMC trends are closely similar, whilst the trends at 8.0 $\mu$m are well separated. This latter difference is attributed to incompleteness in the LMC 8.0 $\mu$m sample.

**Figure 4**

A comparison of 4.5 and 3.6 $\mu$m magnitudes (filled diamonds), and 8.0 and 3.6 $\mu$m magnitudes (open squares) for GBPNe (upper panel) and LMC PNe (lower panel). The sample is much greater in the lower panel, where it appears that there is little evolution of [3.6]-[8.0] colour indices with 3.6 $\mu$m magnitude, and a strict upper limit of [3.6]-[8.0] $\cong$ 3.8 mag upon the colours of the sources. The [4.5]-[3.6] trend is close to that expected for bremsstrahlung and central star emission, and this leads to two closely parallel (and overlapping) sequences of points – those corresponding to nebulae in which bremsstrahlung and line emission is dominant (upper points), and those in which stellar emission is important.

**Figure 5**

Distribution of differing populations of planetary nebulae within the [3.6]-[4.5]/[4.5]-[8.0] colour plane, where the large red disks correspond to the Galactic bulge sources; diamonds to the Acker et al. (1992) Galactic disk sources; squares to the MASH GDPNe of Parker et al. (2006) and Miszalski et al. (2008); and the small blue disks to the LMC PNe – the latter results deriving from data presented in this paper, and photometry published by Hora et al. (2008). It will be noted that the differing categories of sources are similarity scattered throughout this plane.

**Figure 6**

As for Fig. 5, but for the [3.6]-[4.5]/[5.8]-[8.0] (top panel) and [5.8]-[8.0]/[4.5]-[5.8] (bottom panel) colour planes. Note the sharp cut-off in LMC PNe above [5.8]-[8.0] ~ 2 mag (top panel), and how the LMC sources are largely confined to the lower right-hand side of the [5.8]-[8.0]/[4.5]-[5.8] plane (bottom panel).

**Figure 7**

Distribution of differing categories of source with respect to the [3.6]-[4.5] colour index. The lower curve represents the trend for sources in which there are no detections at 5.8 or 8.0 $\mu$m.

**Figure 8**

As for Fig. 7, but for the [4.5]-[8.0] (left panel) and [5.8]-[8.0] (right panel) colour indices. All of these sources are detected in the four IRAC bands.

**Figure 9**

The distribution of LMC PNe (filled diamonds) and GBPNe (open squares) in the NIR colour plane. The large, dashed diamond-shaped region indicates the region of reddened GDPNe, whilst the rectangle corresponds to the regime of de-reddened GDPNe (see Ramos-Larios

& Phillips 2005). The locus for main-sequence stellar colours is taken from Tokunaga (2000). It should be noted that sources to the left-hand side of the plane correspond to nebulae in which emission is dominated by central star continua, and that the disparity between the positions of the bulge and LMC sources indicates a difference in extinction of $\Delta A_V \sim 6$ mag.

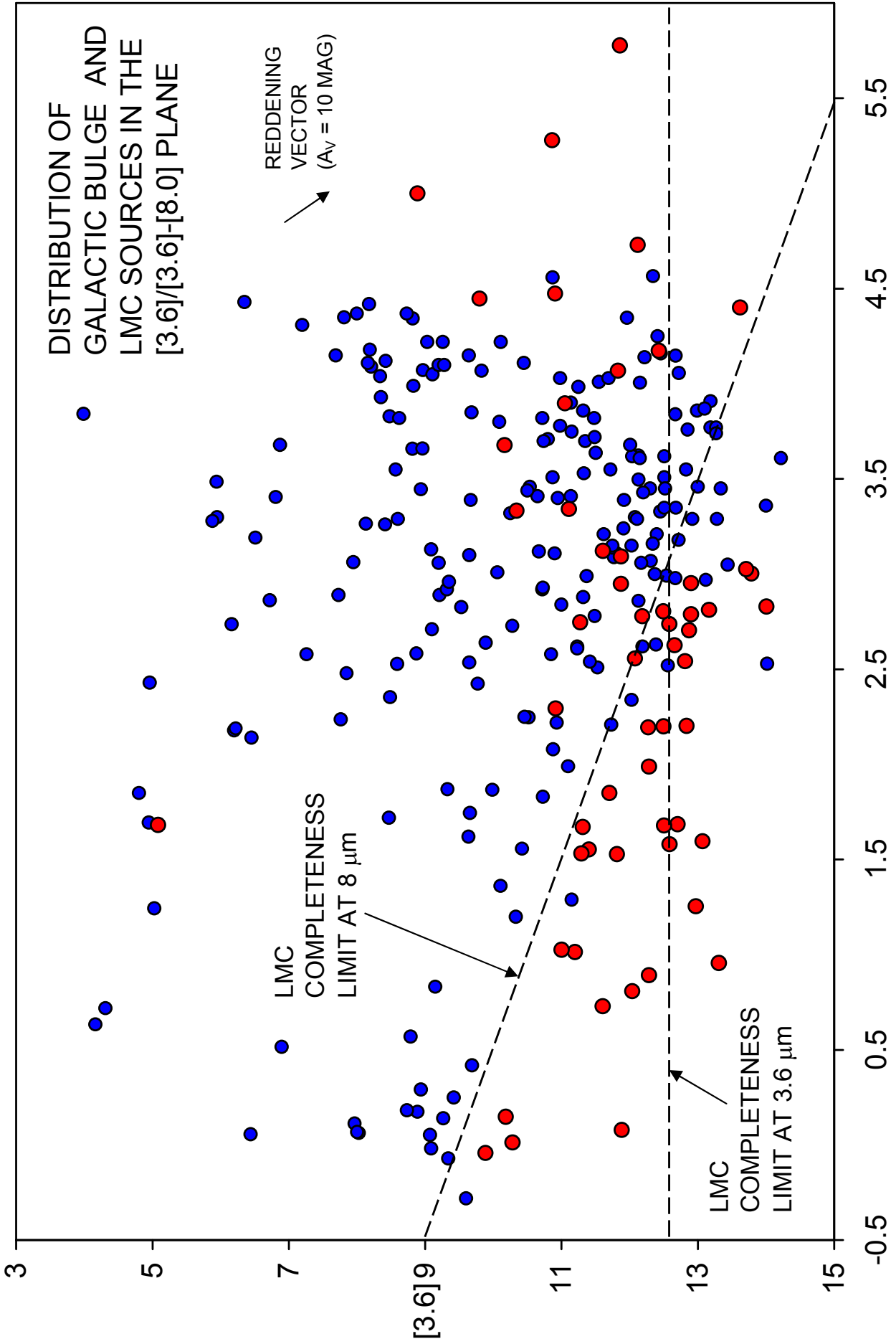

FIGURE 1

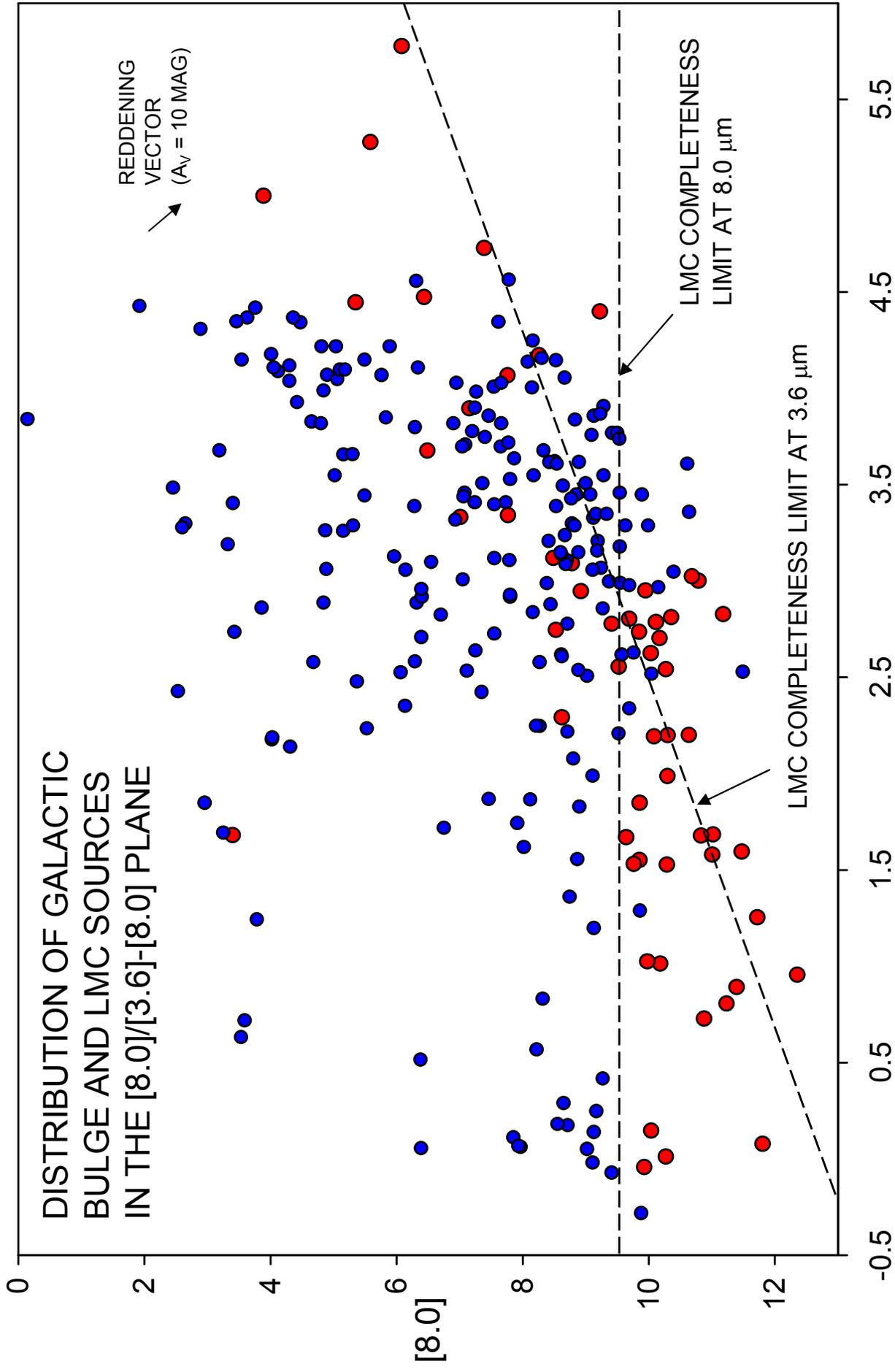

FIGURE 2

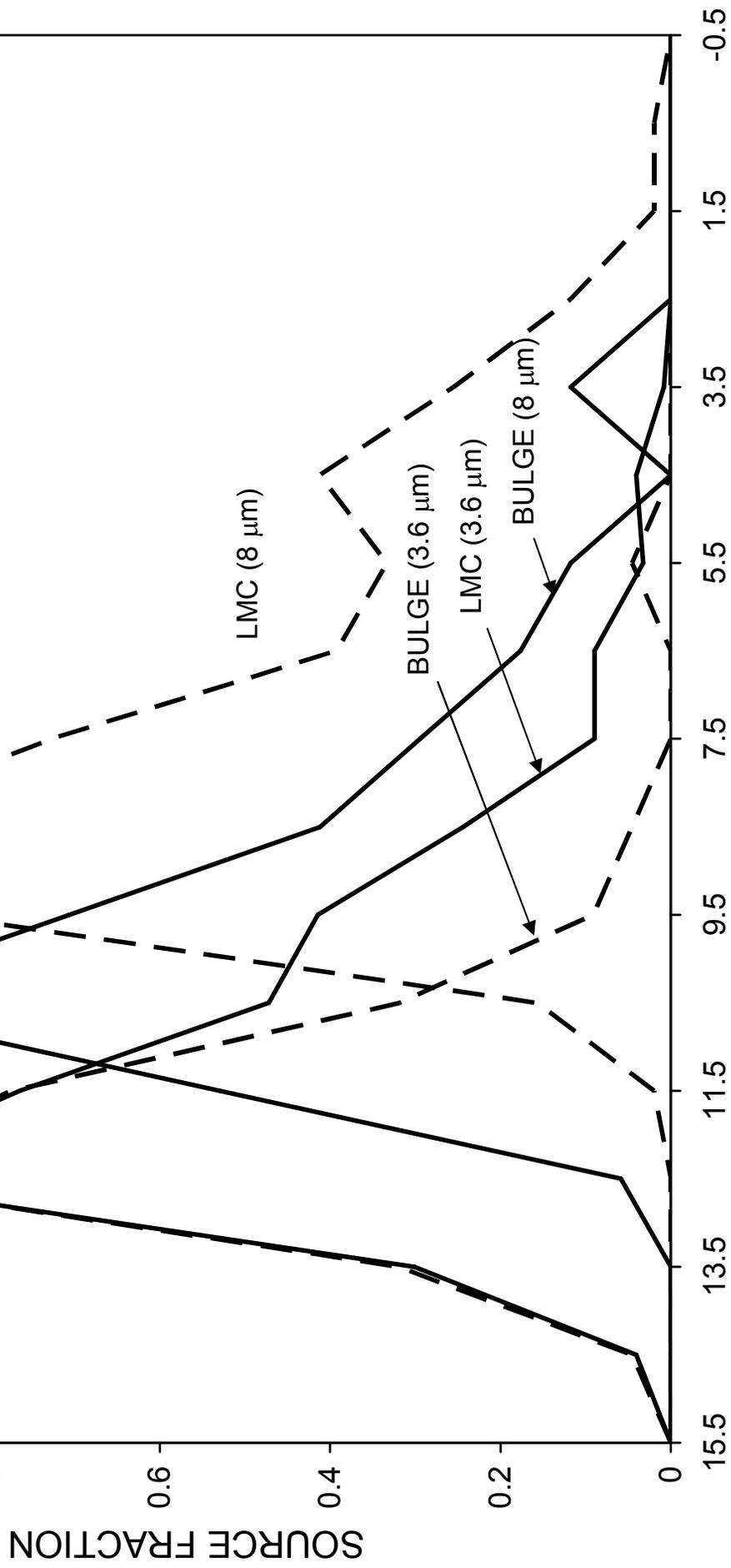

FIGURE 3

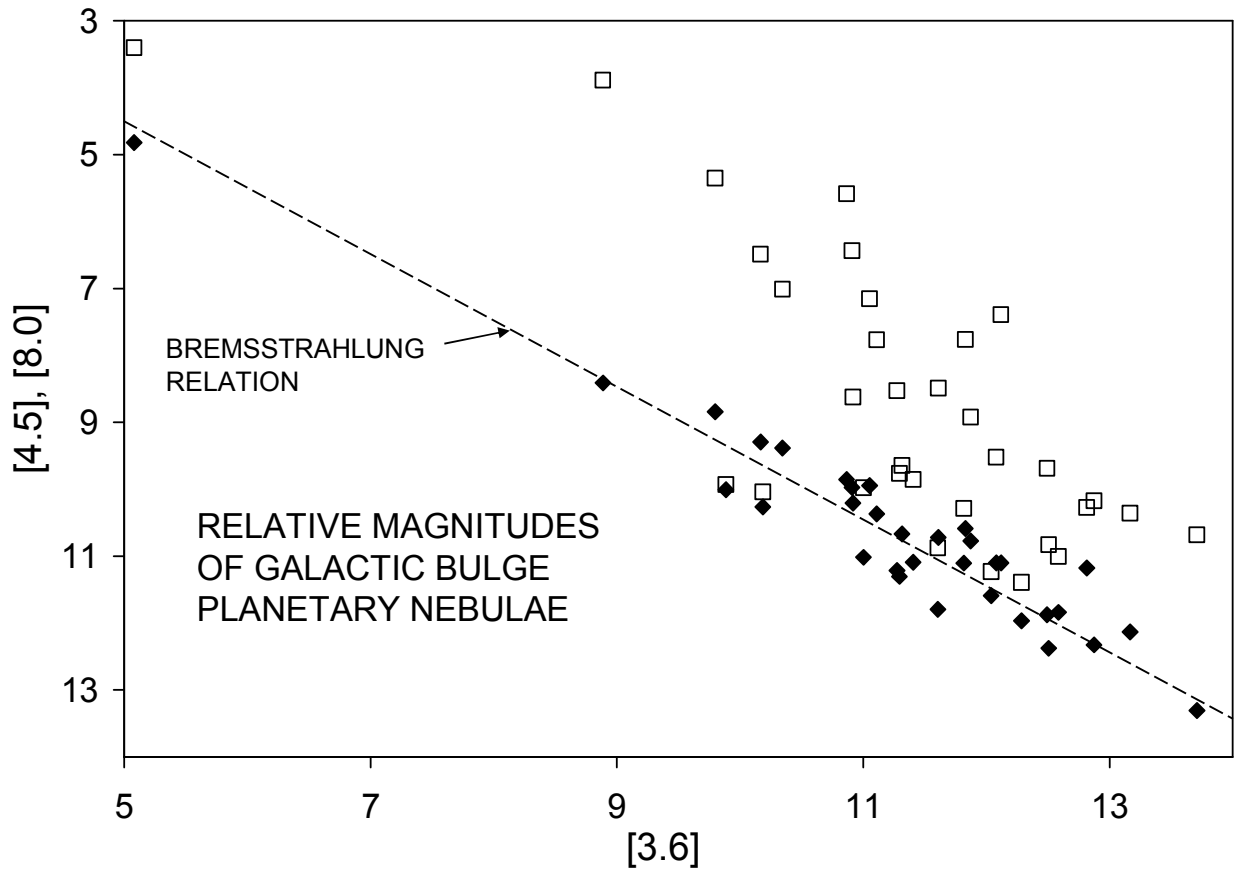
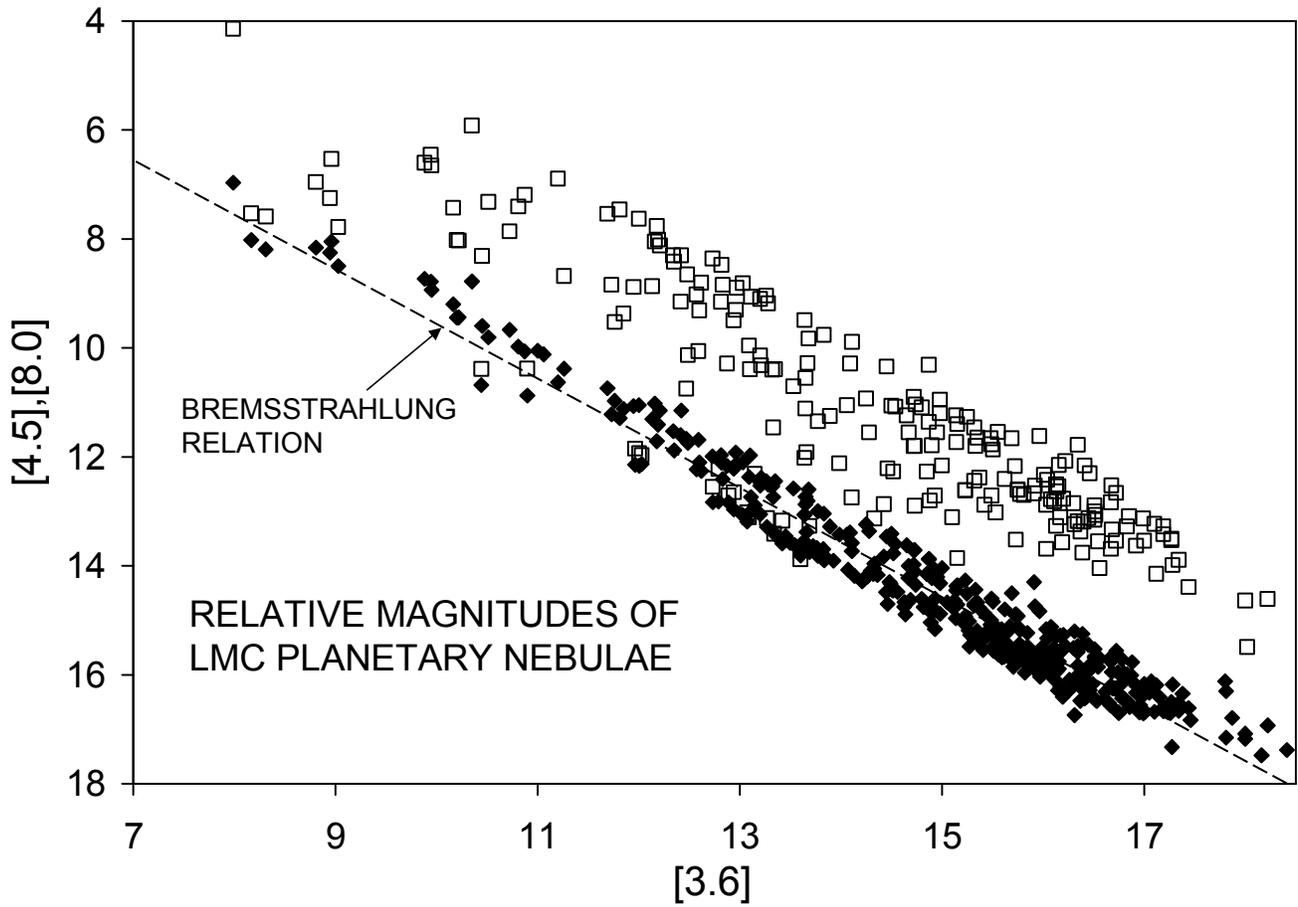

FIGURE 4

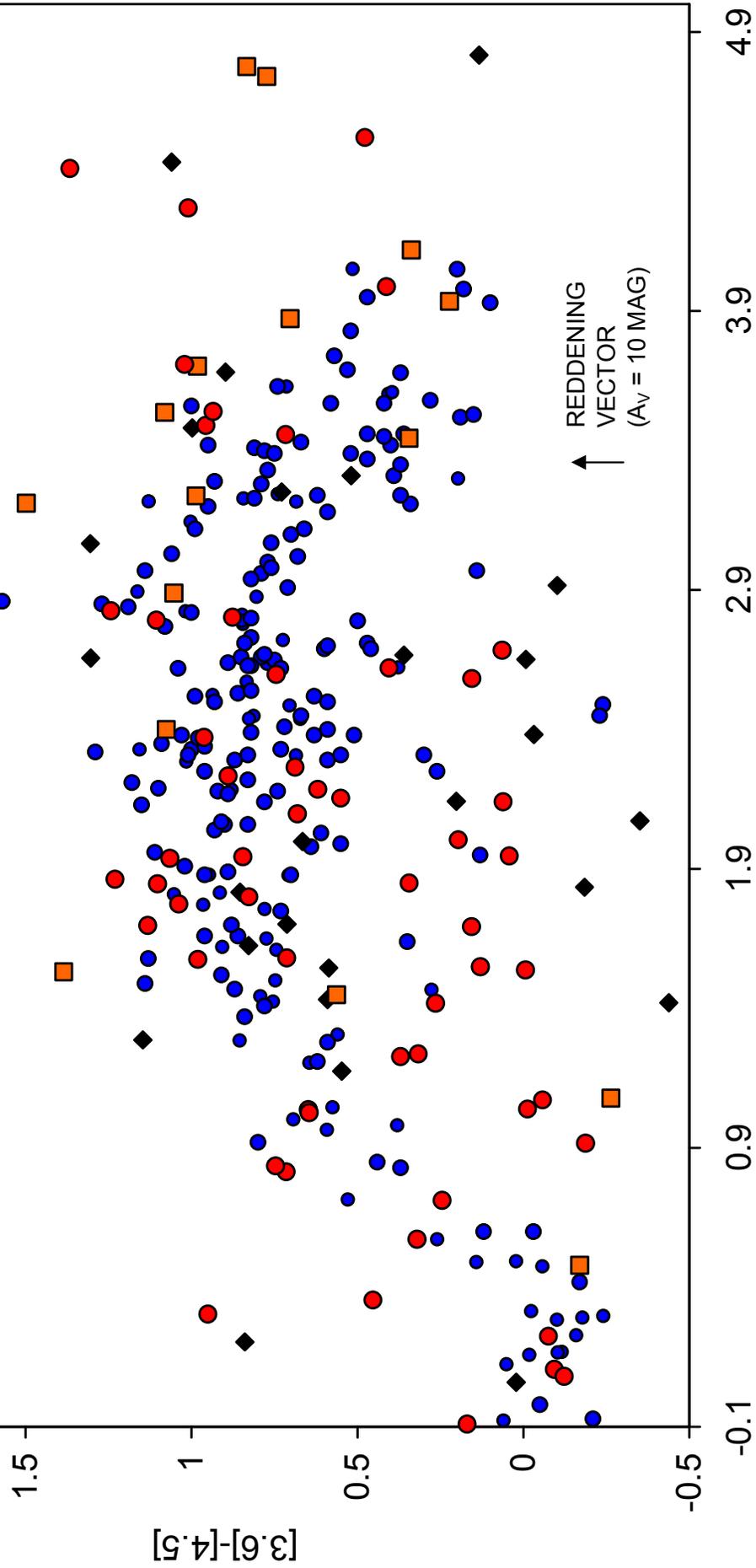

FIGURE 5

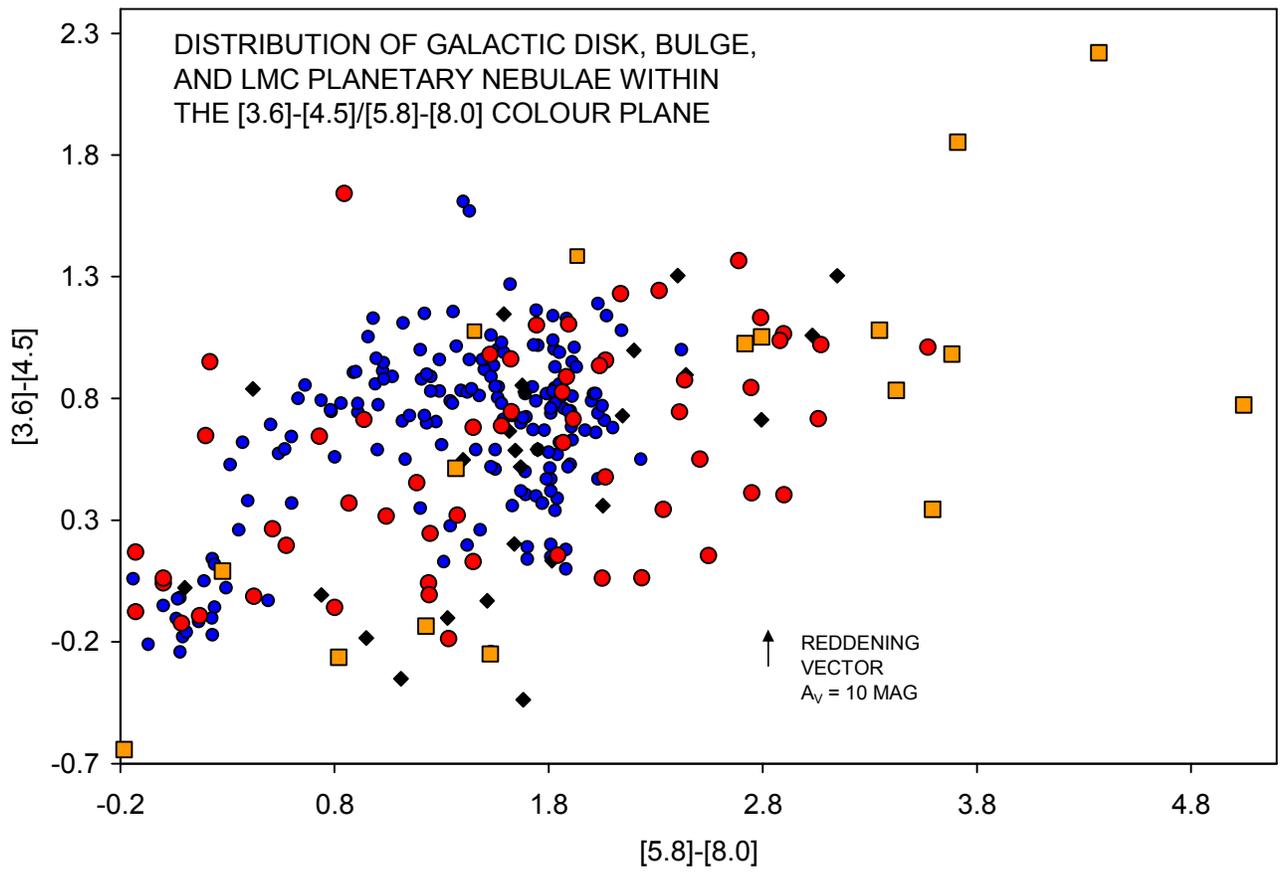
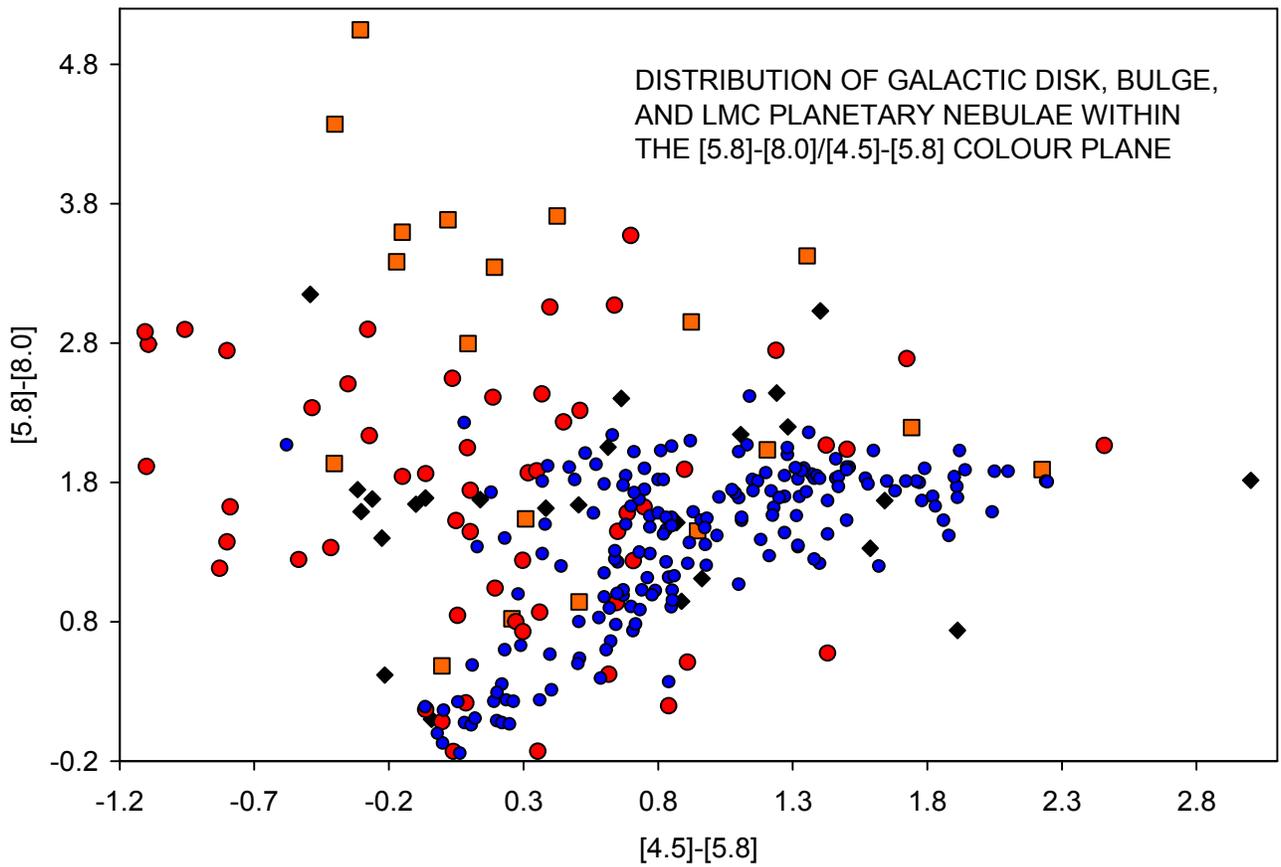

FIGURE 6

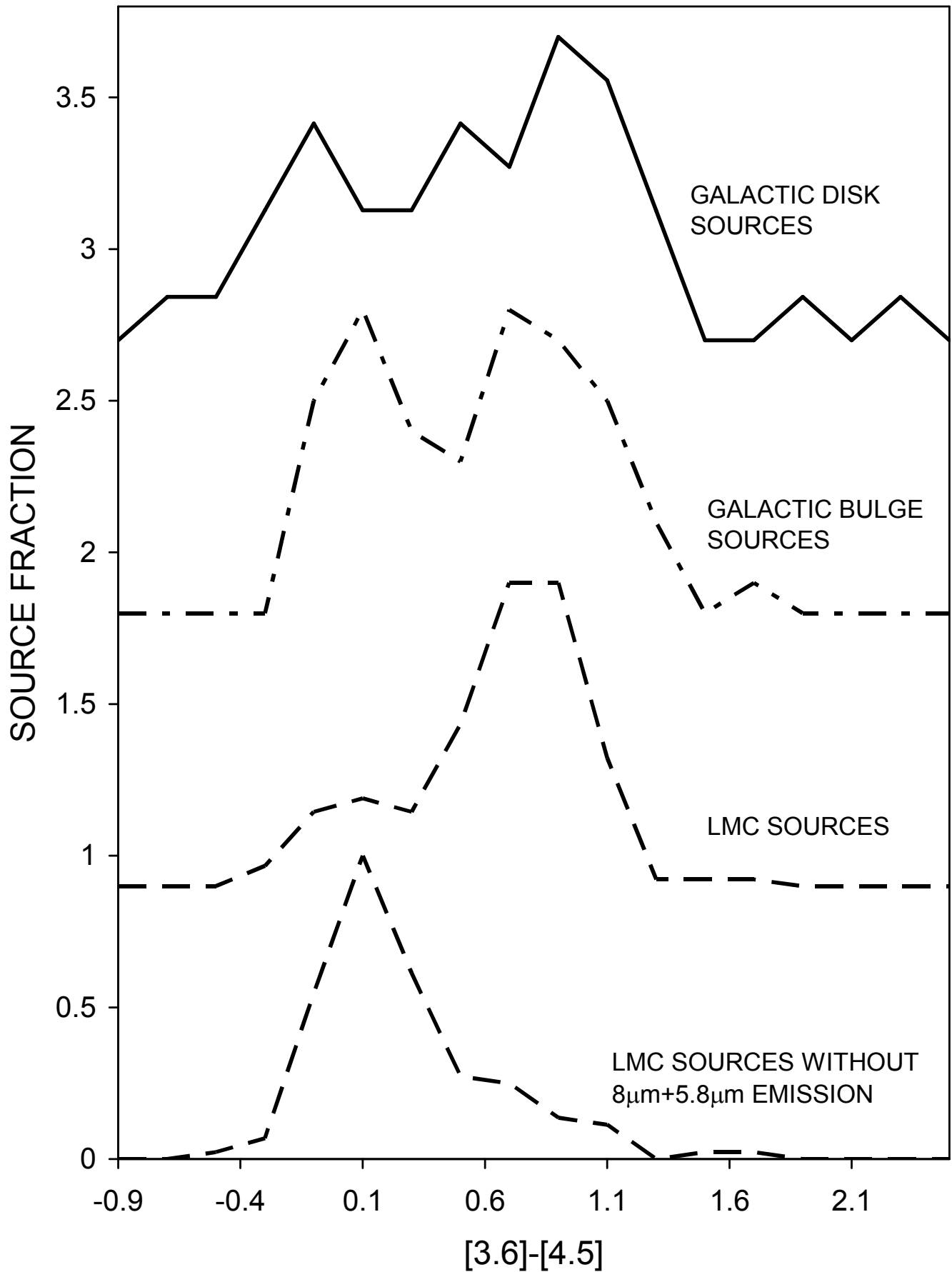

FIGURE 7

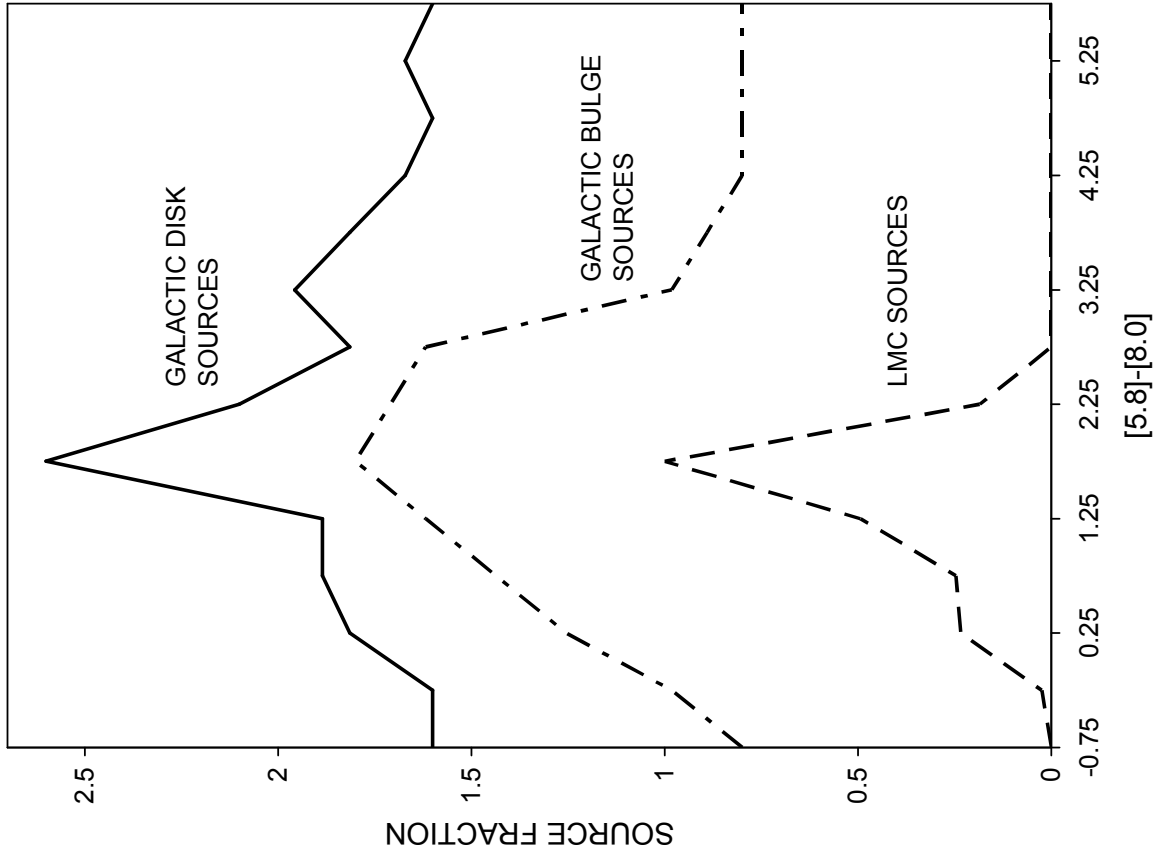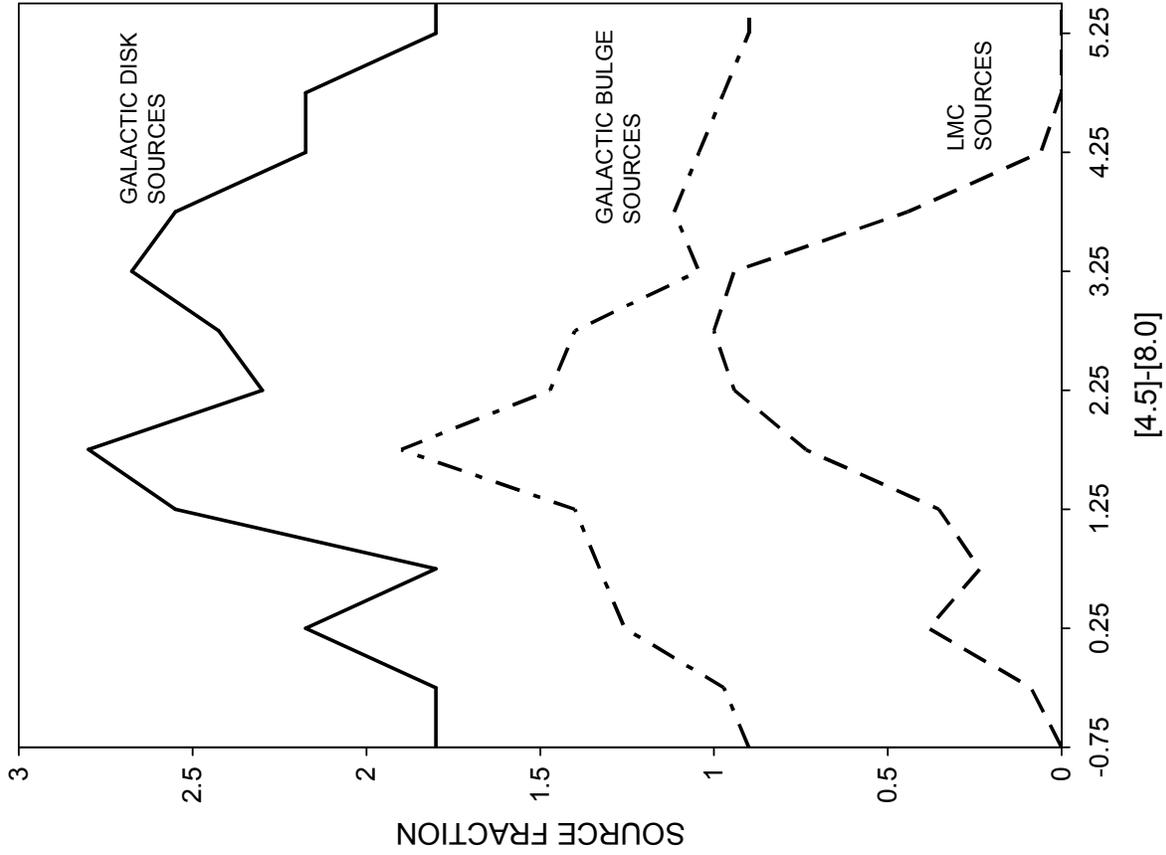

FIGURE 8

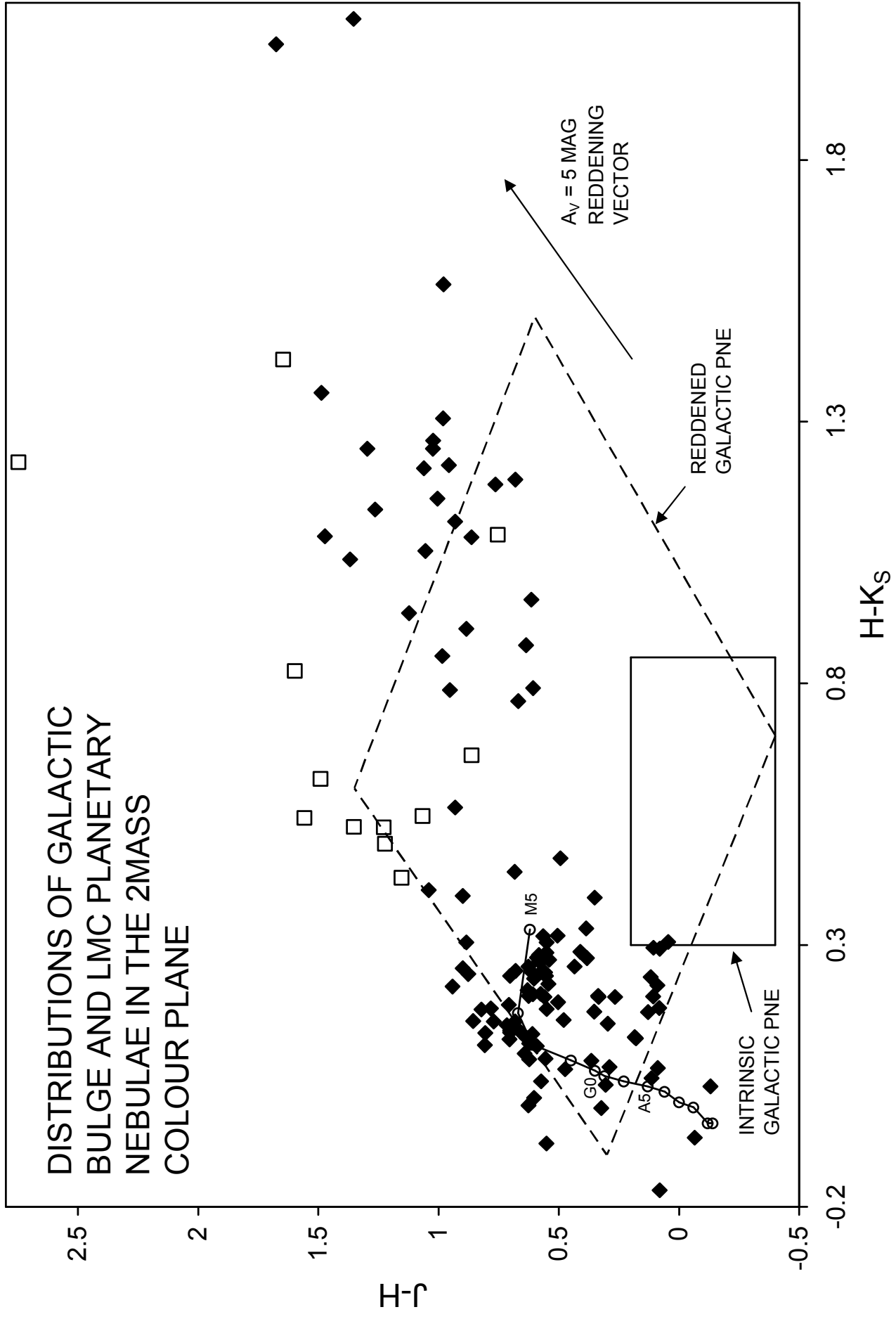

FIGURE 9

58